\documentclass[bibyear]{aa}  
\usepackage{graphicx}
\usepackage{natbib}
\bibpunct{(}{)}{;}{a}{}{,} 
\usepackage{pdflscape}
\usepackage{longtable}
\usepackage[varg]{txfonts}
\usepackage{caption}
\usepackage{subcaption}
\usepackage[caption=false]{subfig}
\usepackage{ltablex,booktabs}

\def\farcs{\hbox{$.\!\!^{\prime\prime}$}}
\def\fdg{\hbox{$.\!\!^\circ$}}
\def\arcsec{\hbox{$^{\prime\prime}$}}
\def\utw{\smash{\rlap{\lower5pt\hbox{$\sim$}}}}
\def\udtw{\smash{\rlap{\lower6pt\hbox{$\approx$}}}}

\begin{document}

\title{Hundreds of new cluster candidates in the VISTA variables in the
  V\'{\i}a L\'actea survey DR1 
      \thanks{Based on observations taken within the ESO VISTA Public Survey
        VVV, Program ID 179.B-2002. The full catalog is only
        available at the CDS via anonymous ftp to 
        cdsarc.u-strasbg.fr (130.79.128.5) or via
        http://cdsarc.u-strasbg.fr/viz-bin/qcat?J/A+A/ .}
    }

   \author{R.~H. Barb\'a\inst{1}\fnmsep\thanks{rbarba@userena.cl},
          A. Roman-Lopes\inst{1},
          J. L. Nilo Castell\'on\inst{1},
          V. Firpo\inst{1},
          D. Minniti \inst{2,3,4,5,6},
          P. Lucas \inst{7}, 
          J.~P. Emerson \inst{8},
          M. Hempel \inst{2,6},
          M. Soto \inst{1,9},
          R.~K. Saito \inst{10}.
          }

   \institute{Departamento de F\'{\i}sica y Astronom\'{\i}a, 
              Universidad de la Serena,
              Av. Juan Cisternas 1200 Norte, La Serena, Chile         
            \and
             Departamento de Ciencias F\'{\i}sicas, Universidad Andres Bello,
             Rep\'ublica 220, Santiago, Chile 
             \and
              Vatican Observatory, Vatican City State V-00120, Italy
             \and
             European Southern Observatory, Vitacura 3107, Santiago, Chile
             \and
             Department of Astrophysical Sciences, Princeton University,
             Princeton NJ 08544-1001, USA 
             \and
             The Milky Way Millennium Nucleus, Av. Vicu\~na Mackenna 4860,
             782-0436 Macul, Santiago, Chile 
             \and
             Centre for Astrophysics Research, Science and Technology
             Research Institute, University of Hertfordshire, Hatfield
             AL10 9AB, UK
             \and
             Astronomy Unit, School of Physics and Astronomy, Queen
             Mary University of London, Mile End Road, London, E1 4NS, UK
             \and
             Space Telescope Science Institute, MD 21117, USA
             \and
             Universidade Federal de Sergipe, Departamento de F\'isica,
             Av. Marechal Rondon s/n, 49100-000, S\~ao Crist\'ov\~ao, SE, 
             Brazil
             }

   \date{Received 22 April 2014; accepted 12 January 2015}

\abstract
  {VISTA variables in the V\'{\i}a L\'actea is an ESO Public survey
  dedicated to scan the bulge and an adjacent portion of the Galactic disk in
  the fourth quadrant using the VISTA telescope and the near-infrared camera
  VIRCAM. One of the leading goals of the VVV survey is to contribute to the
  knowledge of the star cluster population of the Milky Way.} 
  {To improve the census of the Galactic star clusters, we performed a
  systematic and careful scan of the $JHK_{\rm s}$ images of the Galactic plane
  section of the VVV survey.} 
  {Our detection procedure is based on a combination of superficial density
  maps and visual inspection of promising features in the $J-$, $H-$, and
  $K_{\rm s}-$band images. The material examined are VVV $JHK_{\rm S}$
    color-composite images corresponding to the Data Release 1 of VVV.} 
  {We report the discovery of 493 new infrared star cluster candidates. 
  The analysis of the spatial distribution show that the clusters are very
  concentrated in 
  the Galactic plane, presenting some local maxima around the
  position of large star-forming complexes, such as \object{G305},
  \object{RCW~95}, and \object{RCW~106}.  
  The vast majority of the new star cluster candidates are quite compact
  and generally surrounded by bright and/or dark nebulosities. 
  IRAS point sources are associated with 59\% of the sample, while 88\%
  are associated with MSX point sources. GLIMPSE $8\,\mu$m images of the cluster
  candidates show a variety of morphologies, with 292 clusters dominated
  by knotty sources, while 361 clusters show some kind of nebulosity in this 
  wavelength regime.
  Spatial cross-correlation with young stellar objects, masers,
  and extended green-object catalogs suggest that a large sample of
  the new cluster candidates are extremely young. In particular, 104 star
  clusters associated to methanol masers are excellent candidates for
  ongoing massive star formation. Also, there is a special set of sixteen
  cluster candidates that present clear signspot of star-forming activity
  having associated simultaneosly dark nebulae, young 
  stellar objects, extended green objects, and masers.}
  {}

   \keywords{open clusters and associations --
             galaxy: stellar content --
             galaxy: structure --
             infrared: stars --
             surveys
               }

   \titlerunning{Hundreds of new cluster candidates in the VVV survey DR1}
   \authorrunning{Barb\'a et al.}  
   \maketitle
%

\section{Introduction}
Stellar clusters play a fundamental role in the dynamics and chemical
evolution of galaxies. They trace recent star formation history and
display some of the most spectacular scenarios of stellar evolution.
Indeed, knowing how the clustered stellar populations are distributed within
our galaxy is essential in the study of its large scale structure, hence a key
objective of large surveys like \emph{VISTA variables in the V\'{\i}a
  L\'actea} \citep[VVV,][]{2010NewA...15..433M}.
However, newborn stars formed in clustered environments are often embedded in
large or giant molecular clouds that generate heavy obscuration, which hinder
optical  observations of their early development stages, so observations at
infrared wavelengths are preferred \citep{2003ARA&A..41...57L}.

In the particular case of the Milky Way, any study of its global structure
and morphology is difficult because we are looking at it from a
location placed just in the Galactic plane, where the bulk of the gas, dust,
and stars are known to be confined, making the extinction and crowding
extremely complex and high. 
Because of that, and in spite of developments in instrumentation and
observational techniques, a significant fraction of the Milk Way cluster
population remains unknown. 
In fact, it has been estimated that our galaxy should have at least
$\sim 23,000$\footnote{This number is a lower limit for the estimated
  population of star clusters in the Galaxy. \cite{2006A&A...446..121B}
  indicated a total number of open clusters in the range
  $1.8-3.7\times10^5$.} open clusters \citep{2010ARA&A..48..431P}, however
only a few thousand of them have been identified to date.    

Customarily, the detections of new Galactic star clusters are mostly 
based on a visual inspection of recorded photographic and electronic images
(CCDs, IR detectors, etc.) taken both at optical and near-infrared
(NIR) wavelengths, and the development of new homogeneous surveys in the past
decade has allowed the discovery and analysis of many unknown cluster
candidates: e.g., \cite{2003A&A...404..223B}, \cite{2003A&A...400..533D},
\cite{2005ApJ...635..560M}, \cite{2007MNRAS.374..399F},
\cite{2011A&A...532A.131B}, \cite{2014A&A...562A.115S}, etc. 
Particularly relevant to our research is the 
\emph{Two-Micron All-Sky Survey} \citep[2MASS,][]{2006AJ....131.1163S}, 
the prior near-infrared all sky survey, covering the same 
wavelength range as VVV, but different spatial resolution 
($4\farcs0$ vs.$0\farcs7$) and photometric depth ($K_{\rm s}$-mag 15.5 vs. 
18 in VVV).
Also, if we consider that open clusters are gravitationally bound structures
that consist of more or less coeval stellar populations, the detection of
density enhancements, either visually or by means of computational algorithms,
should only be the first step to properly identifying potential new clusters.
Indeed, stellar density enhancements may also be produced by chance alignments
or localized low foreground extinction \citep[e.g.,][]{2002A&A...383..163O,
  2008AN....329..602M, 2010A&A...510A..44M, 2007MNRAS.374..399F}.
In this sense, to confirm that a group of stars are really physically
related and forming a true cluster, some additional criteria such as radial
density profiles (e.g., Gaussian or King), colors, and/or kinematics, are
probably necessary.
Besides that, the Hipparcos catalog (ESA 1997) has provided accurate
positions (as probably GAIA will do in the future), proper motions, and
trigonometric parallaxes for a significant stellar population in the solar
neighborhood.
For example, many OB associations show small velocity dispersions
\citep[e.g.,][]{1996A&AS..118..503T}, which makes them appear as coherent 
structures in the velocity space, and \cite{1999AJ....117..354D} used this  
characteristic to distinguish the membership in the Hipparcos catalog of
nearby OB associations inside a $\sim 1$~kpc radius of the solar neighborhood.  

Here we present the results of our search for new cluster candidates in the
Galactic disk portion covered by the VVV Data Release 1 (DR1) images
\citep[][]{2012A&A...537A.107S}. 
This work is based on superficial density maps and visual inspection of the
associated $J-$, $H-$, and $K_{\rm S}-$band VVV DR1 images.  
As a result almost 500 new Galactic cluster candidates were identified 
in the disk area of the VVV survey.  
To make this catalog, we performed a more systematic search in each 
VVV tile than \cite{2011A&A...532A.131B} using the same set of VVV DR1 
images.
This paper is organized as follow: 
In Section~2, we describe briefly the main characteristics of the VVV. 
In Section~3, we describe the finding methods for the new star cluster
candidates.  
In Section~4, we describe the characteristics of the new star clusters and the
spatial cross-correlation with different infrared and radio source catalogs,
and in the Section~5, we summarize our conclusions.

\begin{figure*}
   \centering
\includegraphics[width=45mm]{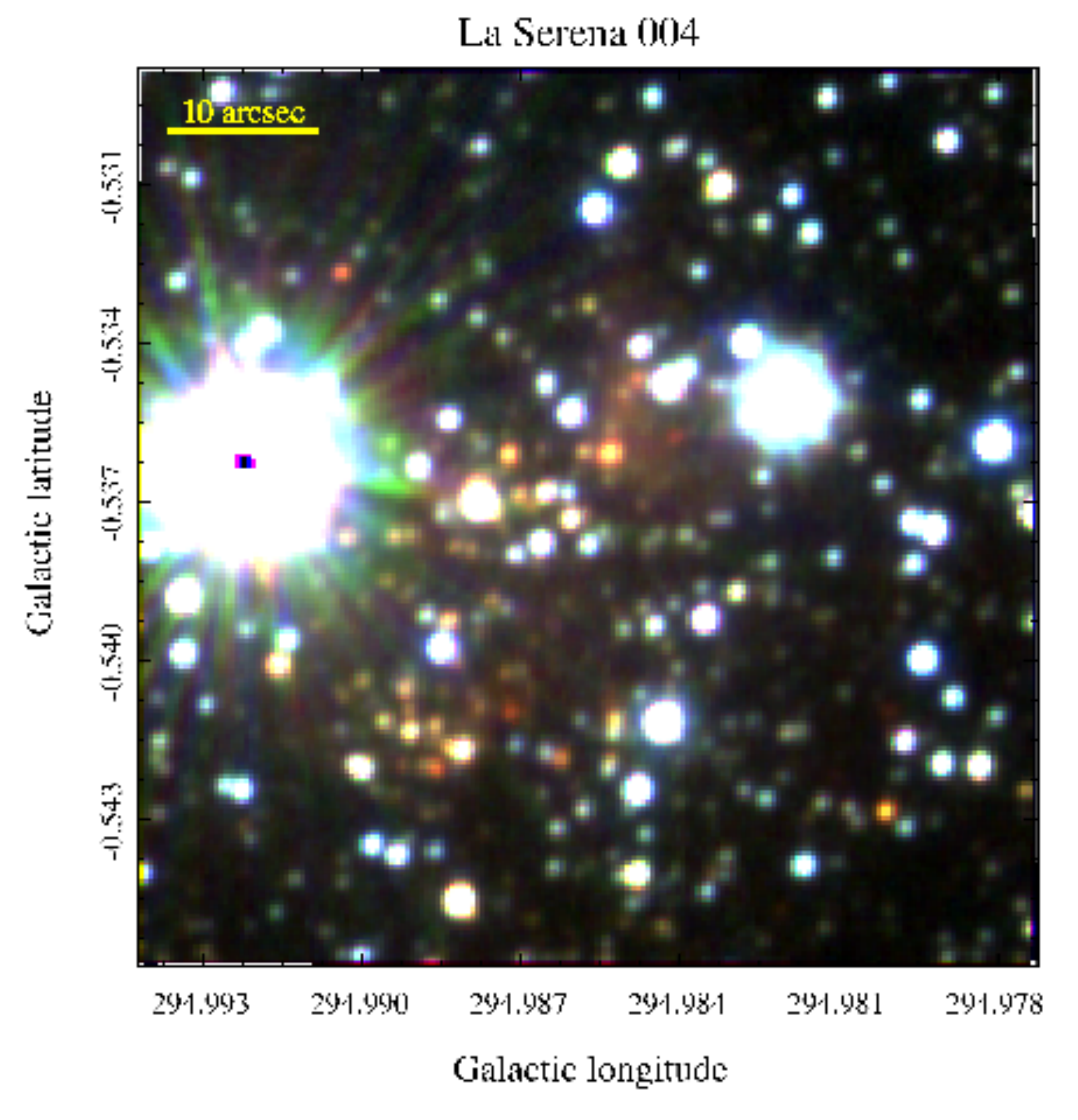}
\includegraphics[width=45mm]{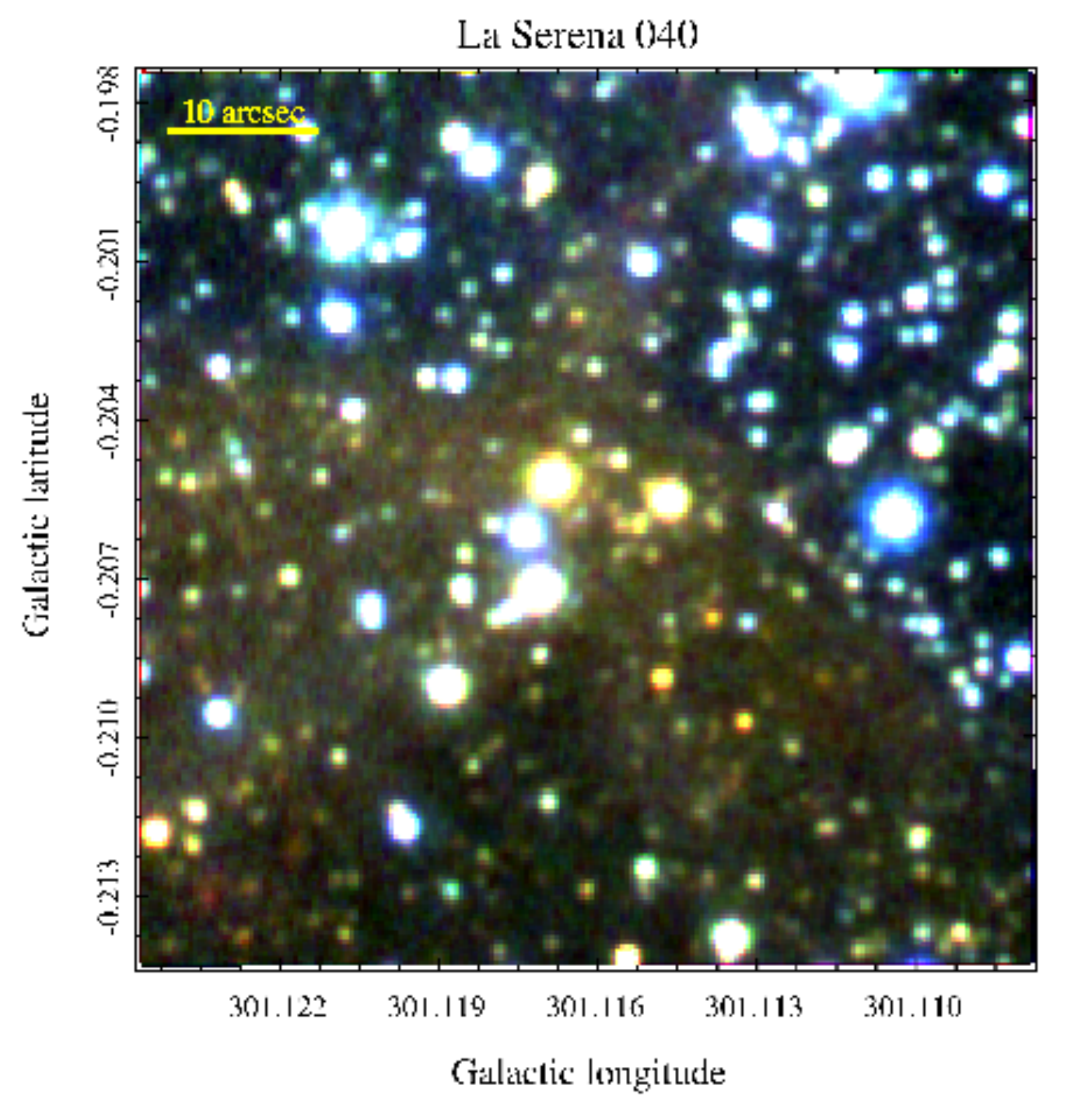}
\includegraphics[width=45mm]{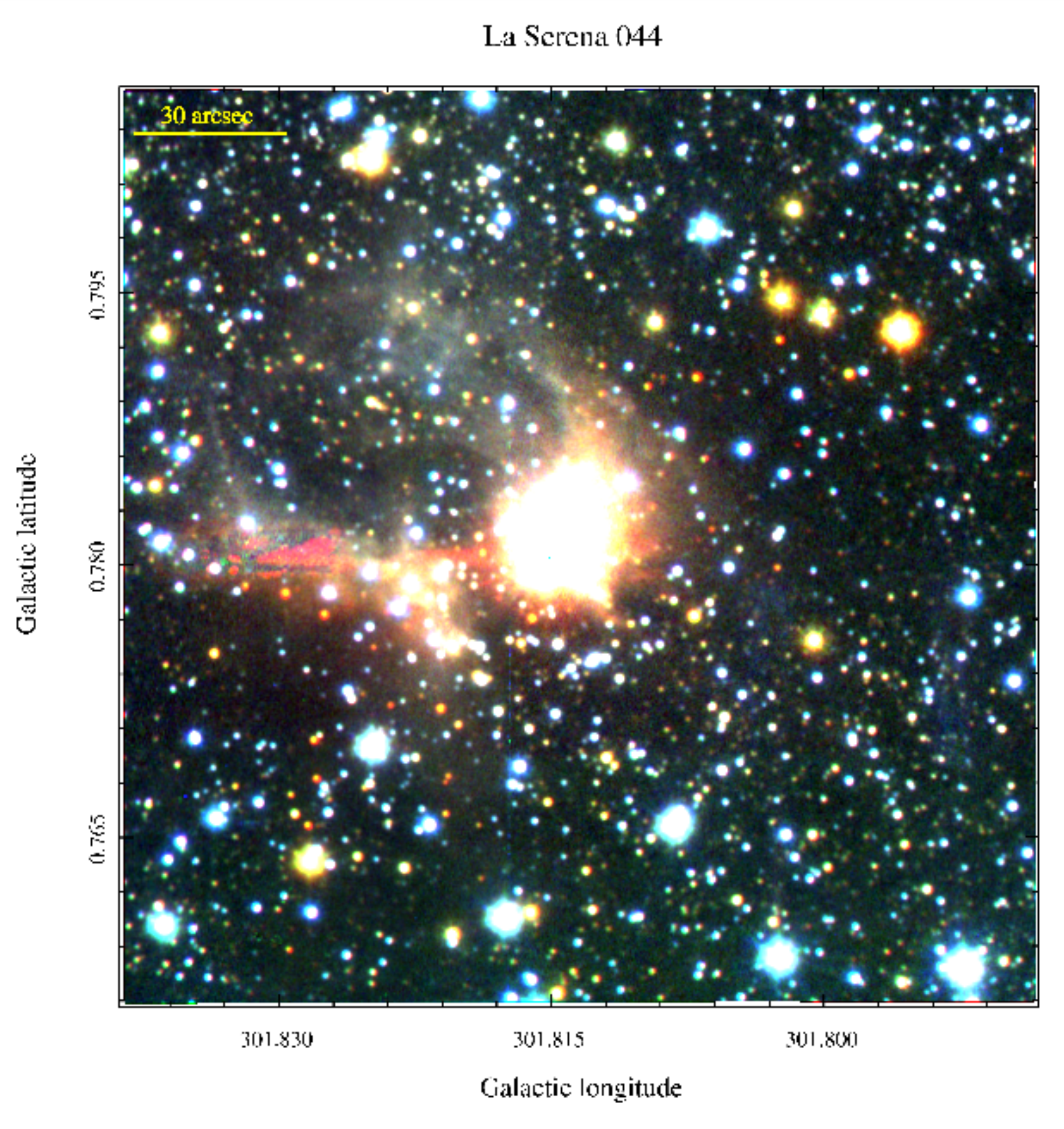}
\includegraphics[width=45mm]{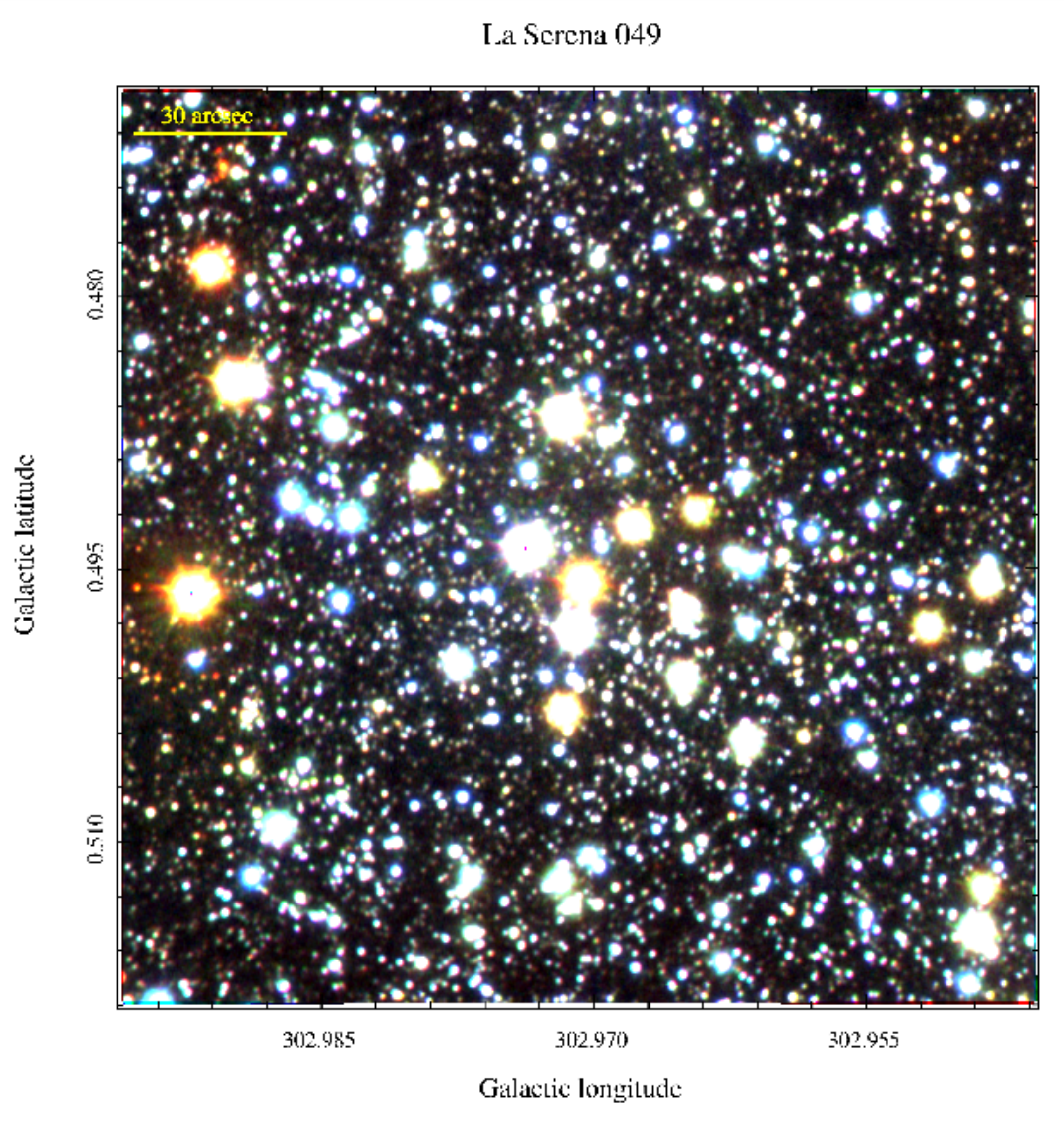}
\includegraphics[width=45mm]{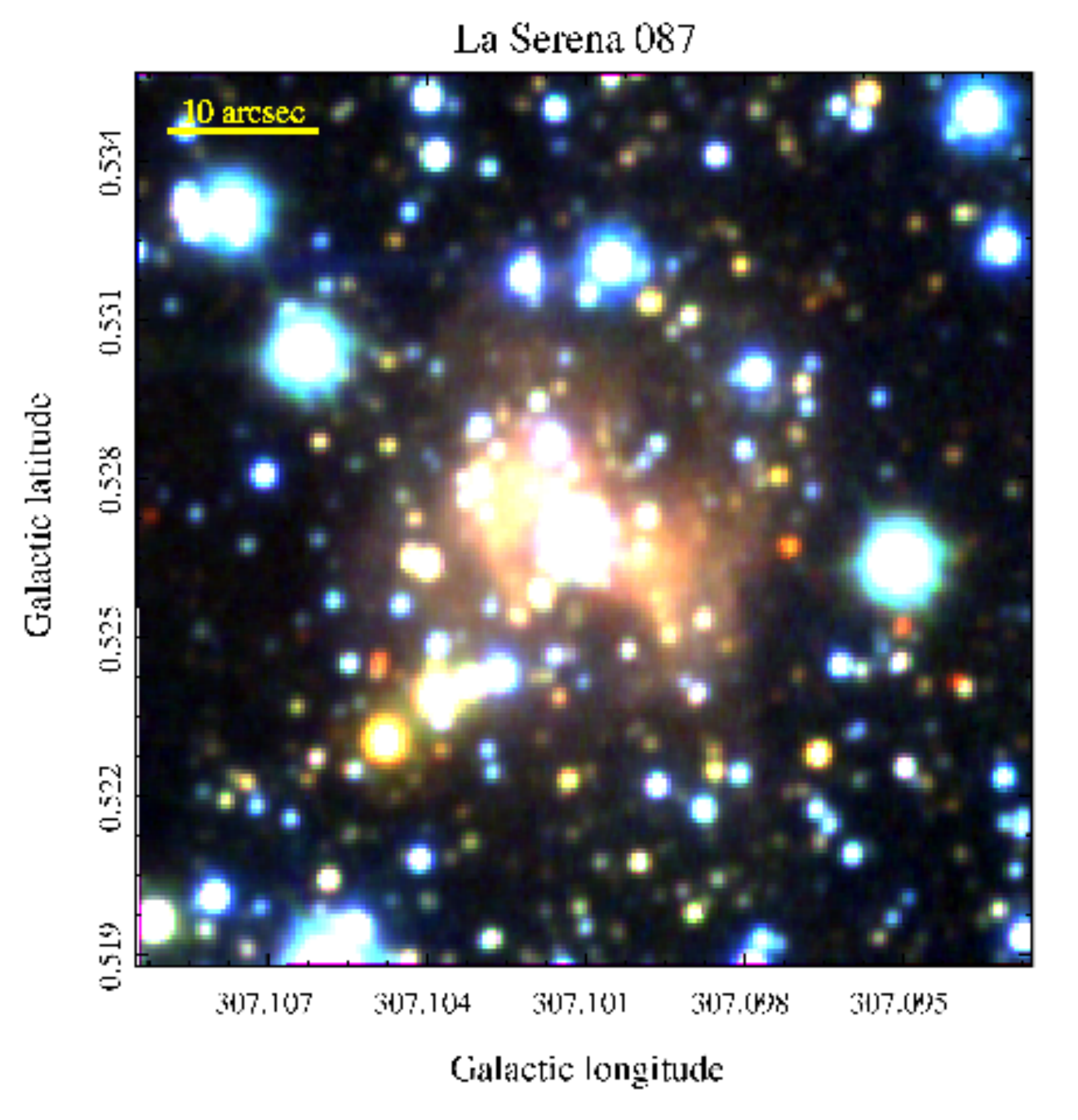}
\includegraphics[width=45mm]{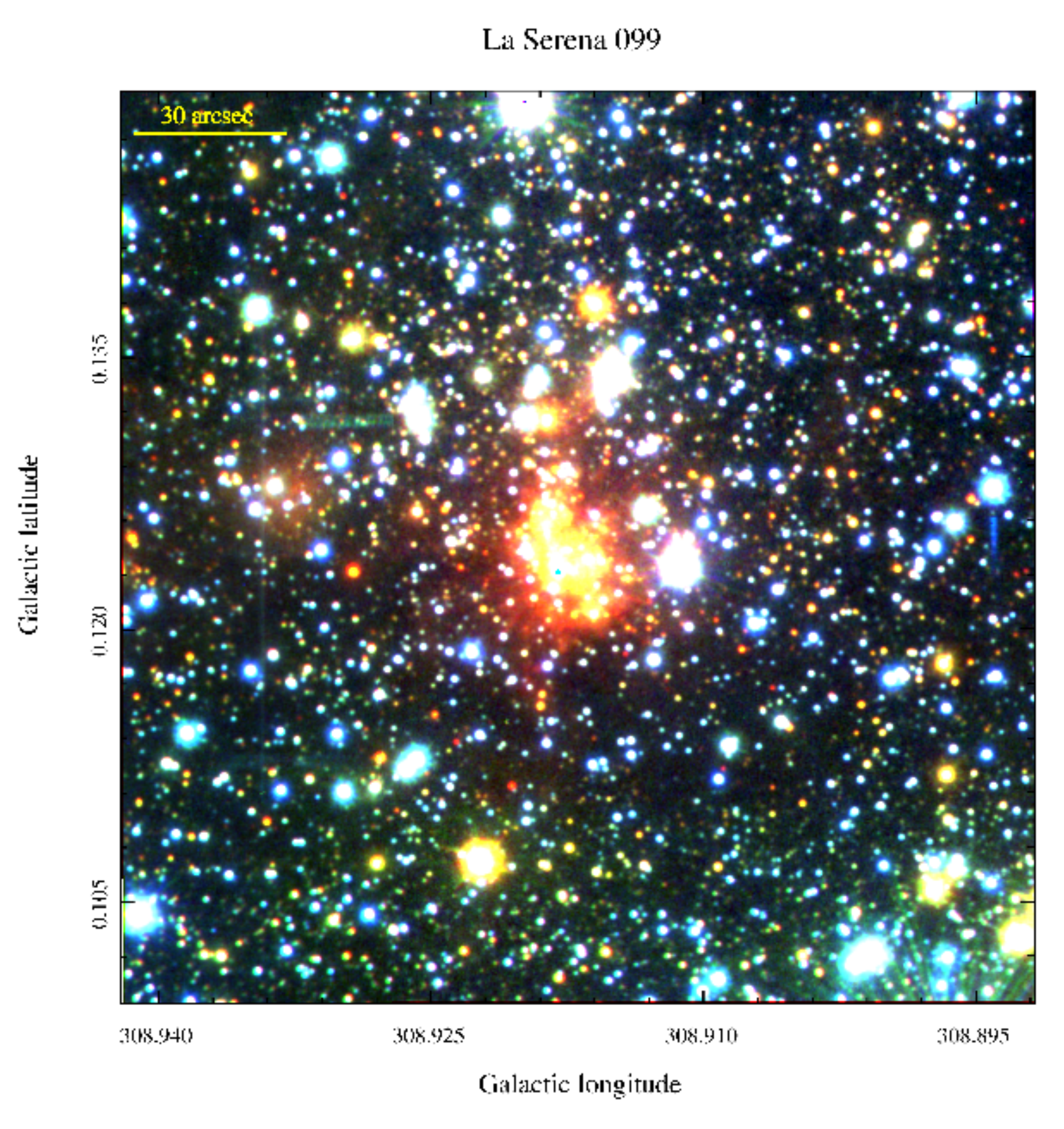}
\includegraphics[width=45mm]{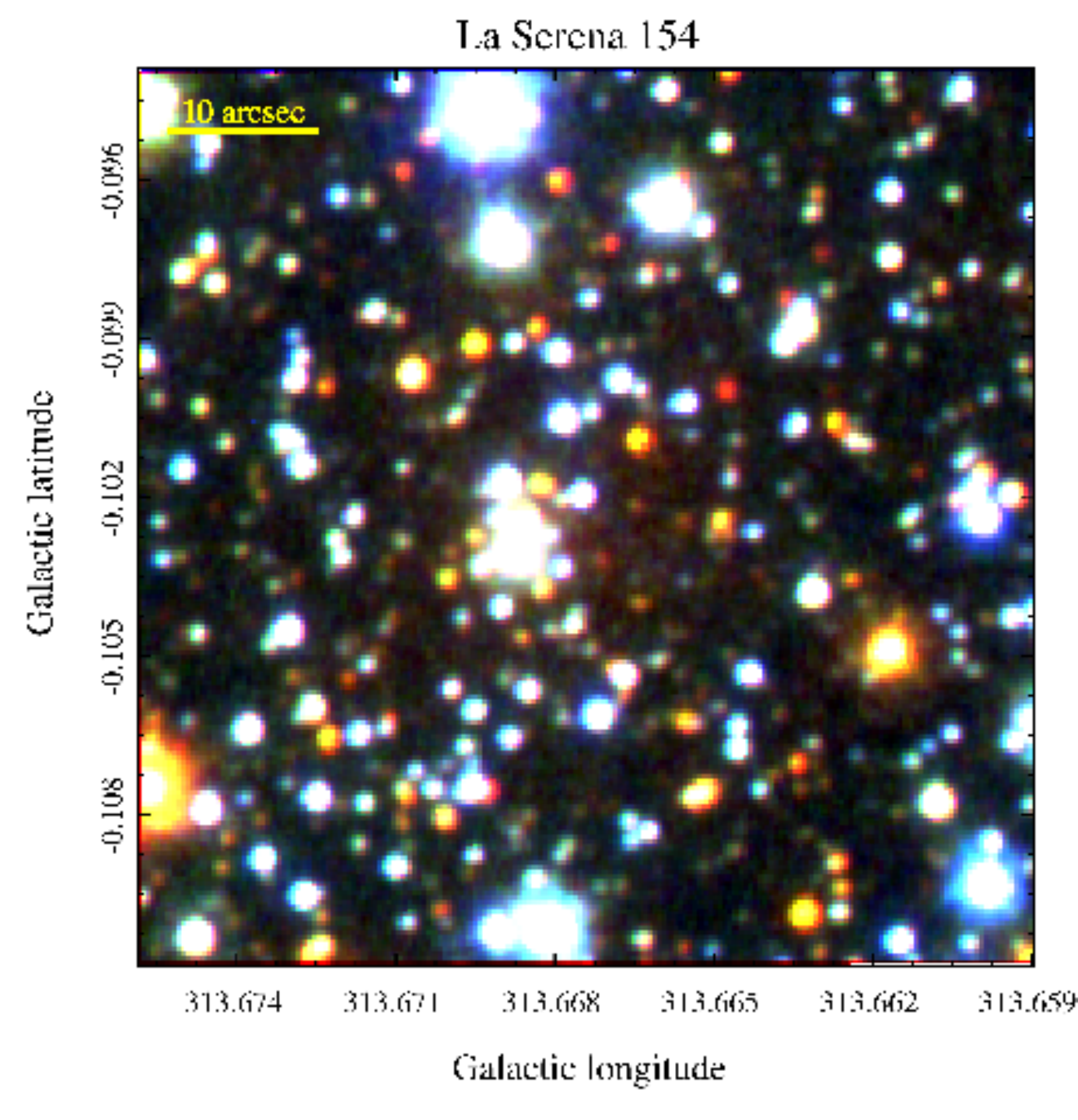}
\includegraphics[width=45mm]{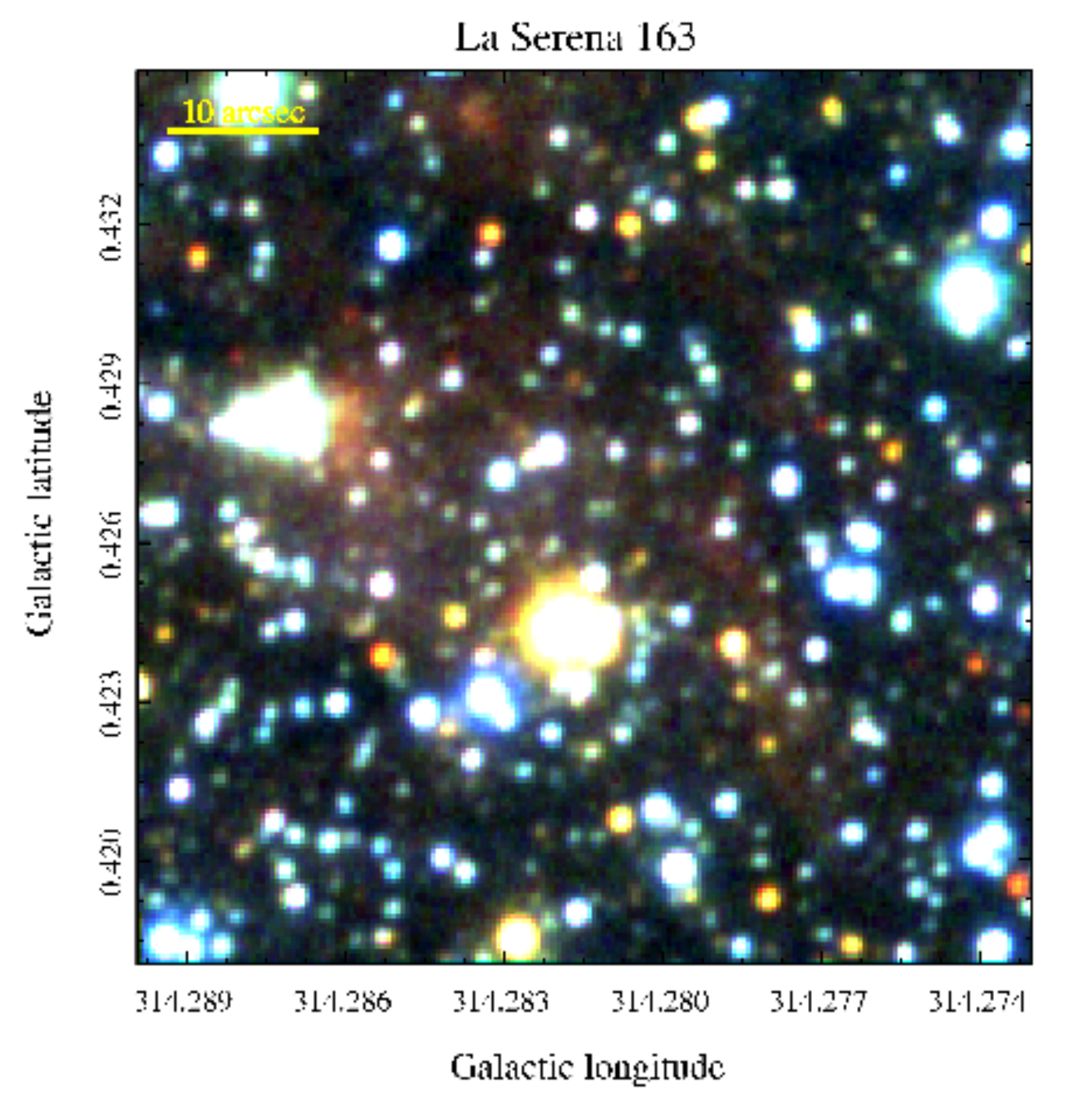}
\includegraphics[width=45mm]{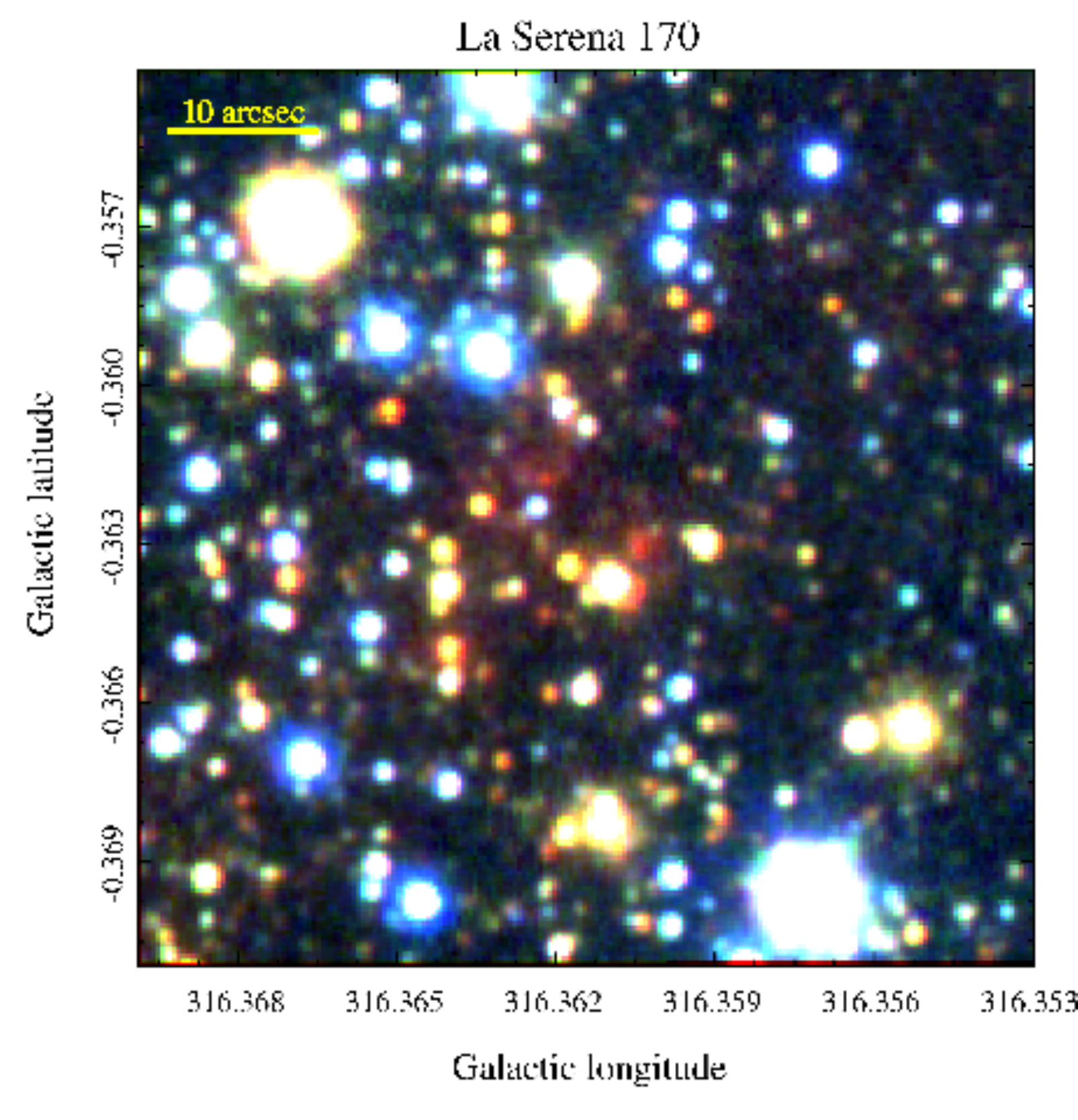}
\includegraphics[width=45mm]{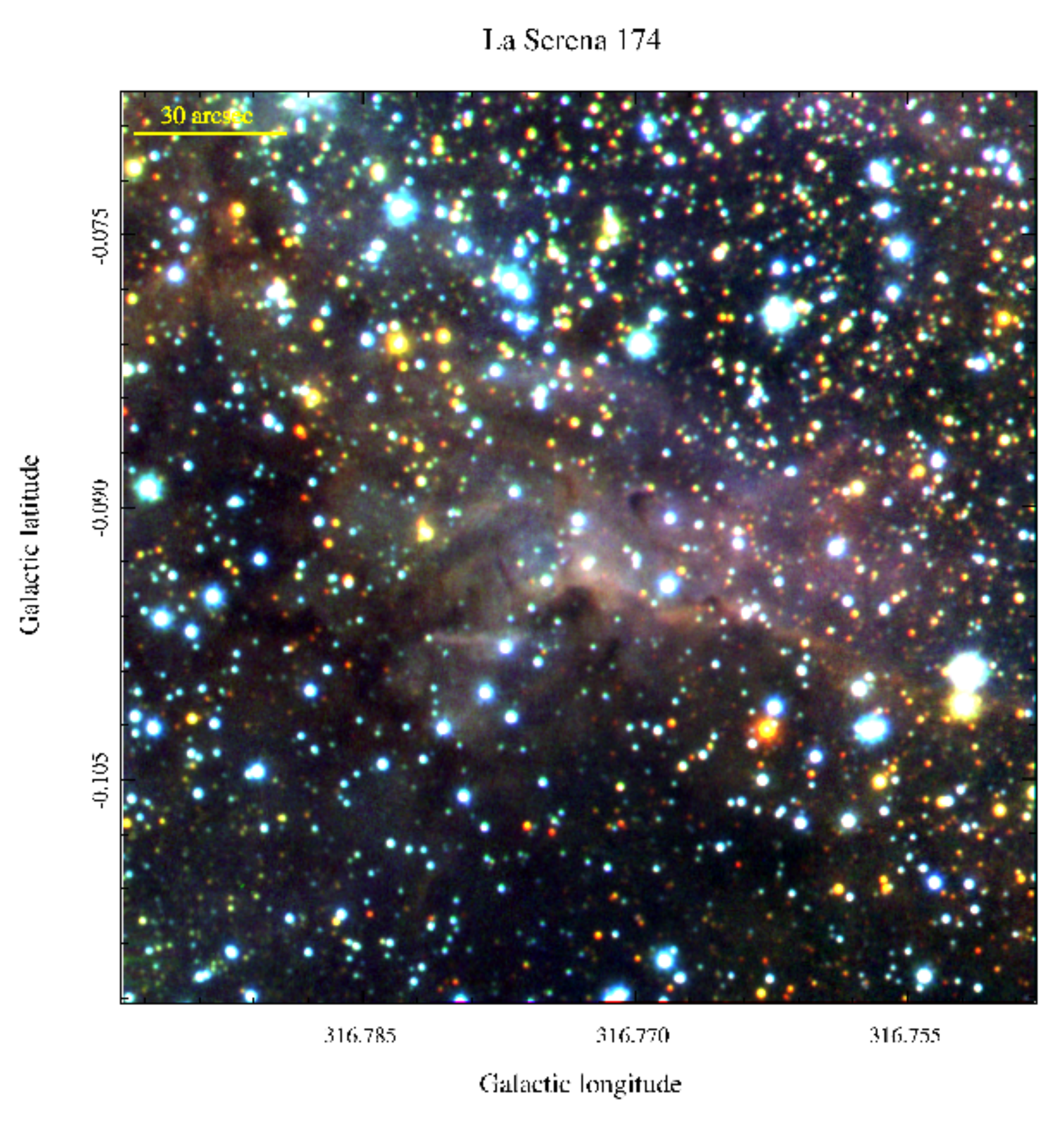}
\includegraphics[width=45mm]{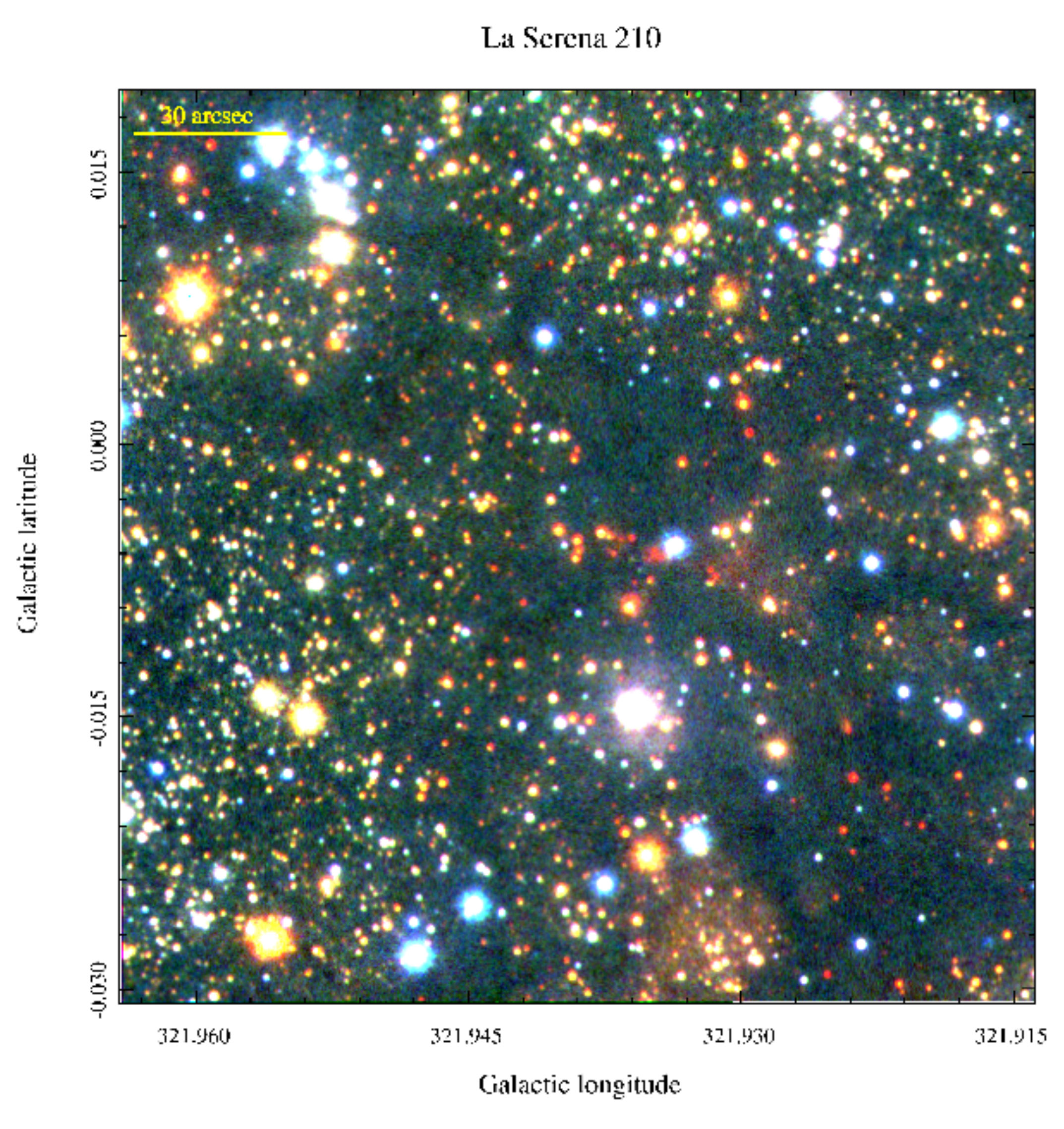}
\includegraphics[width=45mm]{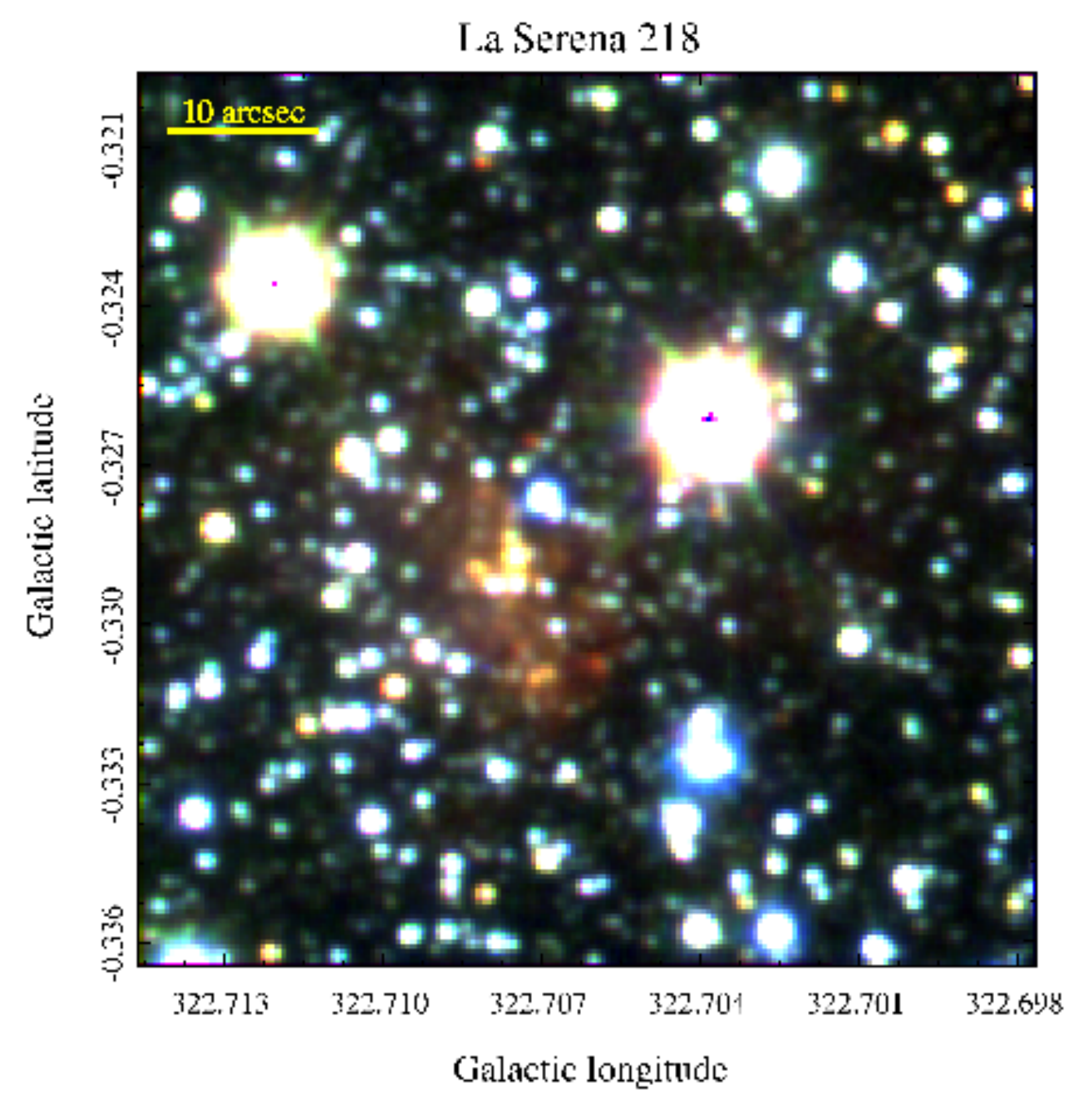}
\includegraphics[width=45mm]{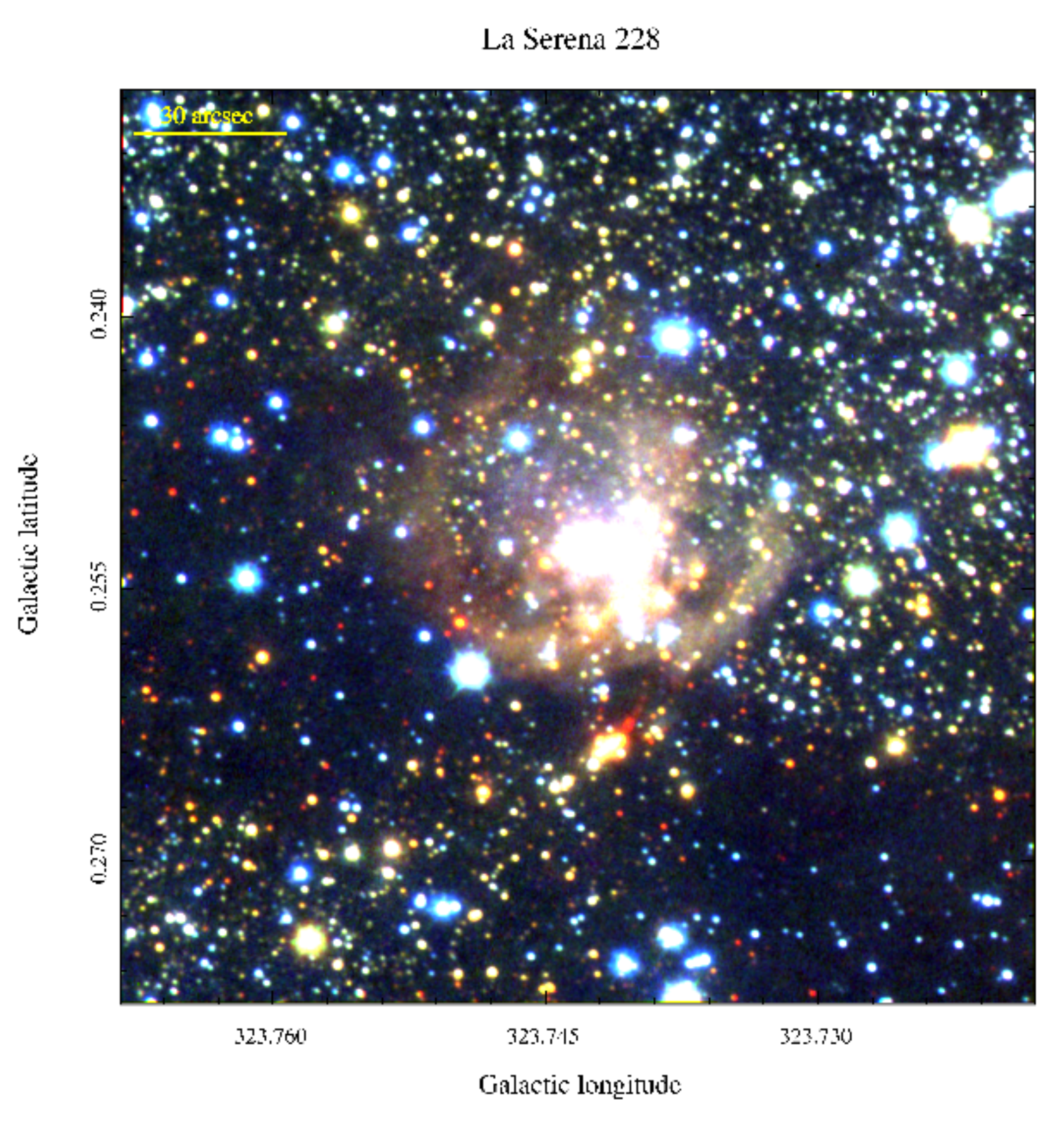}
\includegraphics[width=45mm]{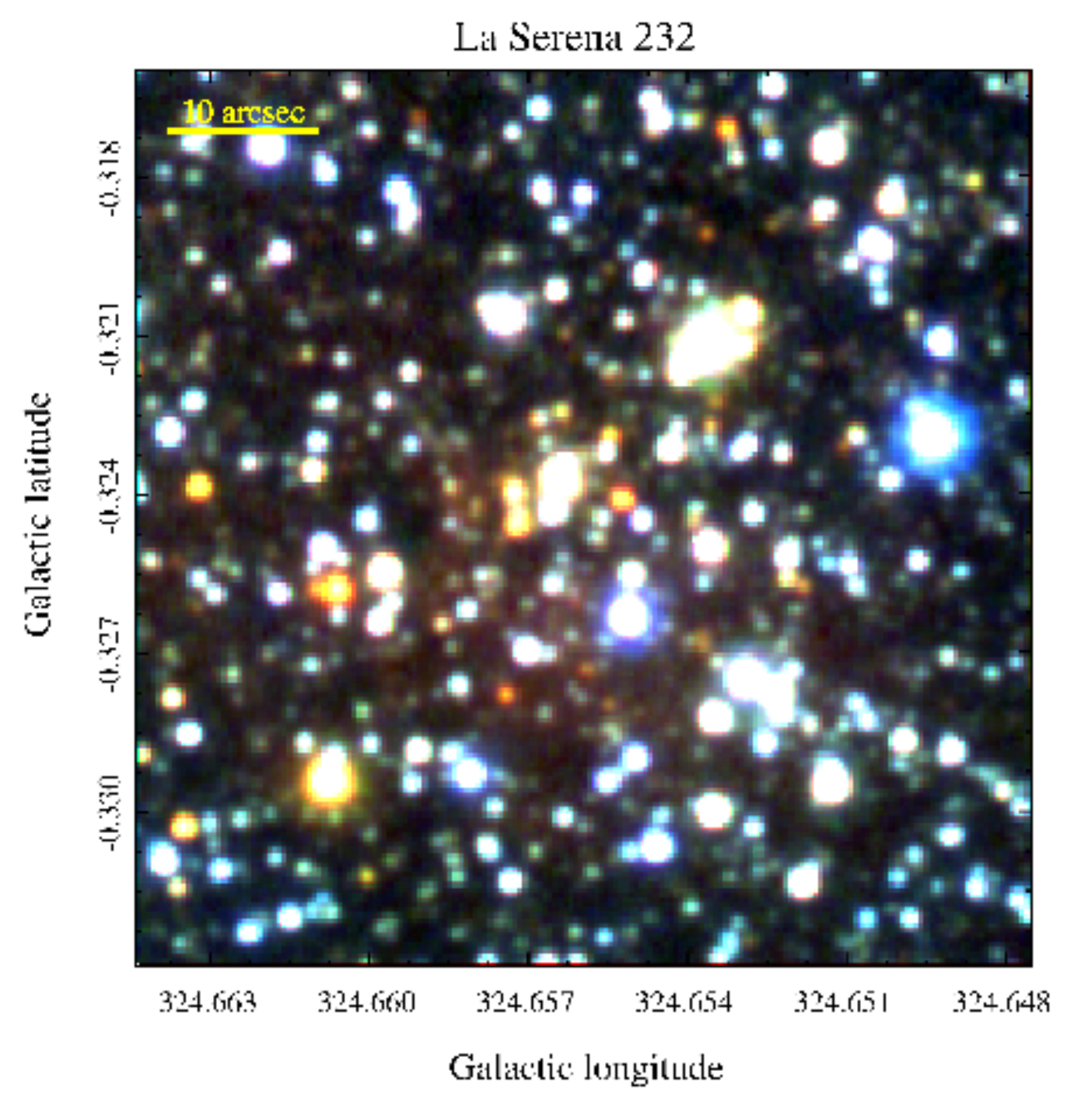}
\includegraphics[width=45mm]{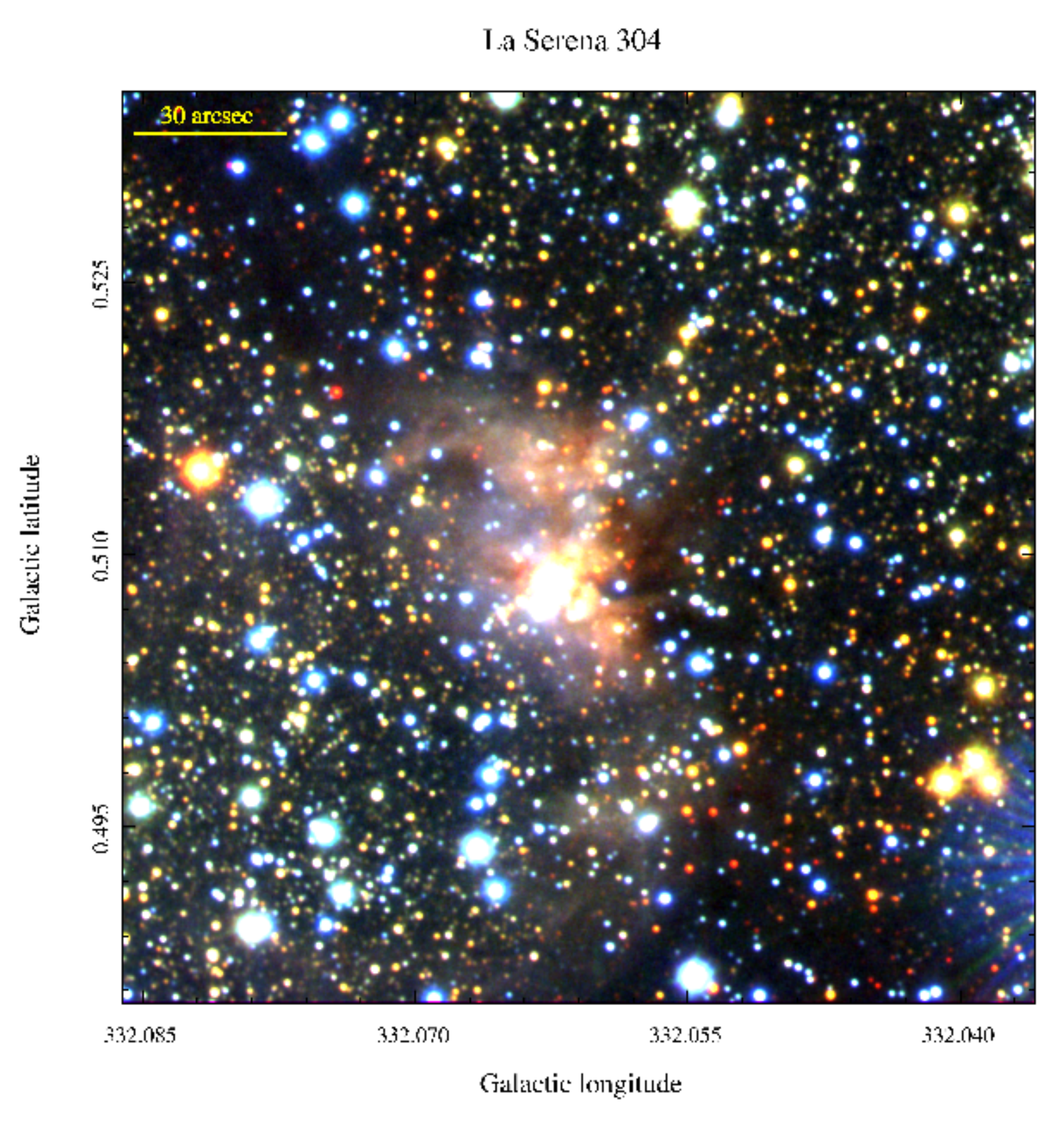}
\includegraphics[width=45mm]{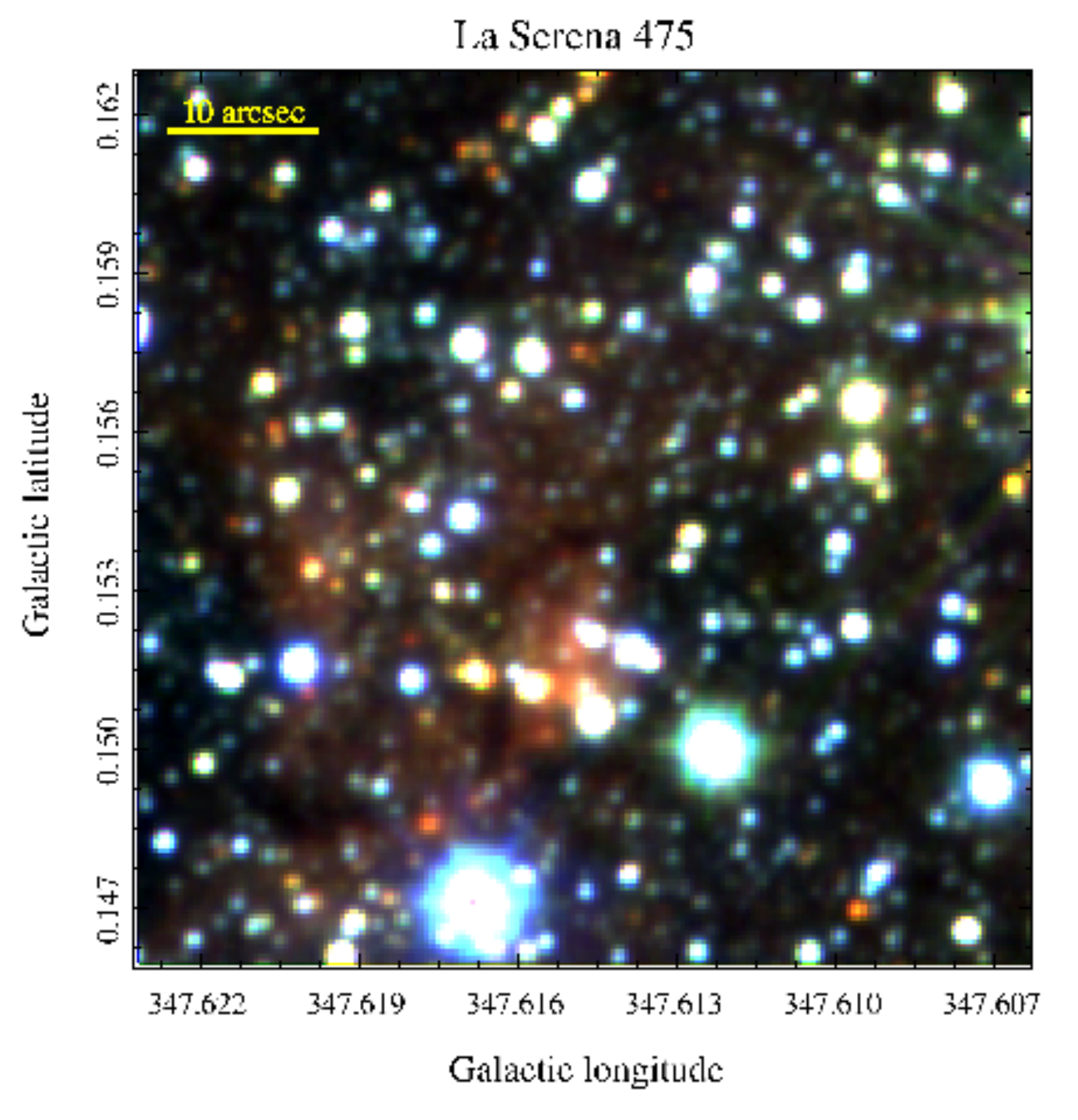}
\caption{RGB color images of a representative sample of the newly discovered
  cluster candidates. The field-of-view for the chart of each cluster is set
  accordingly between $1\times1$ arcmin, $3\times3$ arcmin, and $5\times5$ in
  order to visualize the cluster comfortably. This set of clusters are the
  same showed in Figure\,\ref{spizer_examples}. The complete set of charts for
  the whole catalog is offered in electronic form.} 
  \label{vvv-laserena_examples}
\end{figure*}

\section{Data: VVV tile images}

The VVV survey observations are being carried out in the four-meter Visible and
Infrared Survey Telescope for Astronomy (VISTA), using the VIRCAM \citep[VISTA
InfraRed CAMera;][]{2006SPIE.6269E..30D,2010Msngr.139....2E} located at ESO's 
Cerro Paranal Observatory in Chile.
VIRCAM has a field of view of $1.65^\circ$ by side, but the mosaic of
$4\times4$ Raytheon $2048\times2048$ pixels$^2$ detectors only covers
$0.6$~deg$^2$ in each single exposure (a \emph{pawprint}), with spacings of
40\% and 90\% in the $X$ and $Y$ axes, respectively, and a pixel scale of
$0\farcs34$.
Typically, six VISTA pointings (pawprints) are combined to form a
complete VVV field, a tile.
The VVV survey was planned to observe 348 of these fields in the Galactic
bulge and a portion of the Southern Galactic disk (196 and 152 tiles,
respectively), in five broadband filters $Z$, $Y$, $J$, $H$, $K_{\rm S}$. 
The area covered by the VVV survey overlaps completely with the 2MASS survey,
with part of the Spitzer Space Telescope GLIMPSE survey
\citep{2003PASP..115..953B}, and also complements the area covered by the
UKIDSS survey \citep{2008MNRAS.391..136L}.   

This work is based on the analysis of the 152 $J$, $H$, and $K_{\rm S}$
Galactic plane tile images, which were produced by the Cambridge
Astronomical Survey Unit (CASU) from the scheduled observations of the first
and second years of the survey. 
These images are available as DR1 \citep[][]{2012A&A...537A.107S}.
Further details about the VVV survey, technical description, observing
strategy, and scientific goals are in \cite{2010NewA...15..433M}. 
 
\section{Star cluster candidate finding methods}

Our approach for detecting and selecting each new star cluster
candidate was to look for variations in the stellar density by simple star
counts, helped by the systematic examination of the tile image of each band
and the combined color mosaics displayed in the Aladin virtual observatory
tool \citep[][]{2000A&AS..143...33B}. 
The selected color combination was the usual: $J$, $H$, and $K_{\rm S}$
images combined as blue, green, and red channels, respectively.
The star-count methodology is well known and has been used by several
researchers \citep[e.g.,][]{1995AJ....109.1682L, 1995ApJ...450..201C, 
  2000ApJS..130..381C, 2004MNRAS.353.1025K,2006A&A...449.1033K}.
The method can be refined by using different bin sizes in order to investigate
large-scale structures, as well as smaller scale subclustering
\citep[e.g.,][]{2004MNRAS.353.1025K,2008MNRAS.388..729K}, and/or also by
smoothing the binned data over adjacent bins
\citep[e.g.,][]{2009A&A...497..703K}.    

The star-count technique was applied to the Galactic disk portion of the VVV
DR1 $K_{\rm S}$-band images. 
The stars were detected after using of the task {\sc DAOFIND} (Package
  {\sc Daophot}) in IRAF\footnote{IRAF is distributed by the National Optical 
Astronomy Observatories, which are operated by the Association of Universities
for Research in Astronomy, Inc., under cooperative agreement with the National
Science Foundation.}.  
The limiting magnitudes are those for point sources presenting minimum peak
count values, of $3\sigma$ above the local sky, which in general correspond to
$J$, $H$, and $K_{\rm S}$ magnitude values of about 19.5, 18.8, and 18,
respectively.
Of course, such values must be considered as mean values because the numbers
depend critically on the position of the sources relative to zones of
extended emission, as well as to the heavily crowded zones like those in the
Galactic plane.
It consists in dividing each tile into small bins of equal size and
determining the number of stars in each bin. 
As result, those bins presenting values greater than $3\sigma$ above what was 
measured in a given control region were considered as locations of potential
candidate clusters, deserving further ``by eye'' examination.
The binning sizes were chosen carefully such that the number of objects per
bin was neither too small (prohibiting a meaningful analysis) nor too large 
(hiding existing features). 

We used tipical bin sizes varying from $30$ pixels (about $10\arcsec$) to
about $90$ pixels (about $30\arcsec$), which were determined empirically on
known cluster candidates in the H\,{\sc ii} regions \object{RCW~106} and
\object{RCW~95}, which were previously studied in the past by one of us (ARL). 

As a quantitative example of the procedure applied in the detection of a local
maximum in the superficial density maps on the $K_{\rm s}$-band VVV images,
for a ``small'' cluster candidate like La~Serena 004 (with an estimated 
radius smaller than $30\arcsec$), the binning size that enabled us to detect
it with a appropriate contrast against the surrounding field stars was
$5\arcsec$. 
On the other hand, for a ``large'' cluster candidate like La~Serena 012, the
best results were achieved when we applied the biggest binning size of
$30\arcsec$. 
Of course, we must point out that using this binning range possibly led to a
cluster candidate detection rate biased toward small-to-medium sized cluster
candidates, with most of the larger and sparse clusters being hard to detect.
The control regions used in determining of the mean background to be
subtracted from the statistics for each cluster candidate were computed from
concentric rings with a mean radius equivalent to two to three times the 
observed estimated size of the respective cluster candidate.
We also tested the use of adjacent control fields placed at several arc
minutes (typically 2-5) from the center of the cluster candidate, finding no
significative differences from each method.

Finally, surface density maps were then compared with the color-composite
$JHK_{\rm S}$ images, which in turn were used to properly evaluate and identify
(by visual inspection of the local concentrations of stars) those features
related to candidate clusters. 
Many cluster candidates have affected the star counts owing to the presence of
saturated stars. 
We must take into account that in the VVV images, the saturation limit in the
$K_{\rm S}$-band images is about magnitude 12.5.    
Additionally, we visually identified some compact stellar groups mostly
associated with bright and/or dark nebulosities.  
If such a close association was confirmed, the cluster candidate was evaluated
for overdensities and included in the catalog.  
Also, during the search, we discovered some concentrations of saturated
bright sources in the VVV images, but their appearance resembles a cluster 
of red giant or supergiant stars, similar in morphology to some 
clusters discovered in the 2MASS survey  \citep[cf.][]{2006ApJ...643.1166F,
  2009A&A...498..109C}.    
We included some of these groups after considering the 2MASS data for the
saturated stars, for example, the star cluster candidate labeled as
La~Serena 109. 
We checked carefully that each stellar group candidate had been not discovered
previously, and any previously known neighbor cluster should be located as at
the minimum distance of one arcminute.

Catalogs of star clusters and star cluster candidates used for the check are  
\cite{2003A&A...404..223B}, \cite{2003A&A...400..533D},
\cite{2005ApJ...635..560M}, \cite{2007MNRAS.374..399F}, and
\cite{2011A&A...532A.131B}.
Recently, \cite{2014A&A...562A.115S} have published a new catalog with star
cluster candidates, using the same set of the Galactic disk VVV DR1 images and
also including the bulge area. 
These authors report 88 new star cluster candidates and 39 star formation
location candidates detected by a fit of a mixture model of Gaussian densities
and background noise, and the expectation maximization algorithm to prefiltered 
near-infrared survey stellar catalog data.  

Interestingly, this independent search of cluster candidates is a very good
alternative to checking the reliability of our methods. 
In the VVV tiles of the Galactic disk, \cite{2014A&A...562A.115S} detect 82
star cluster candidates, 34 star formation location candidates, and 24
infrared nebulae of unknown nature. 
We performed a cross-match with a radius of $60\arcsec$ between our catalog and
\cite{2014A&A...562A.115S} sources, obtaining a very good agreement. 
We ``recover'' 111 sources from the complete list of 140 objects in the
\cite{2014A&A...562A.115S}, which are 105 sources inside the $15\arcsec$ 
cross-matching radius. 
The ``recovered'' objects from the \cite{2014A&A...562A.115S} correspond to 
68 cluster candidates (83\% of Solin et al. list), 25 star forming region
candidates (73\%), and 18 infrared nebulae (75\%).

\begin{figure*}[t!]
   \centering
\includegraphics[width=18cm]{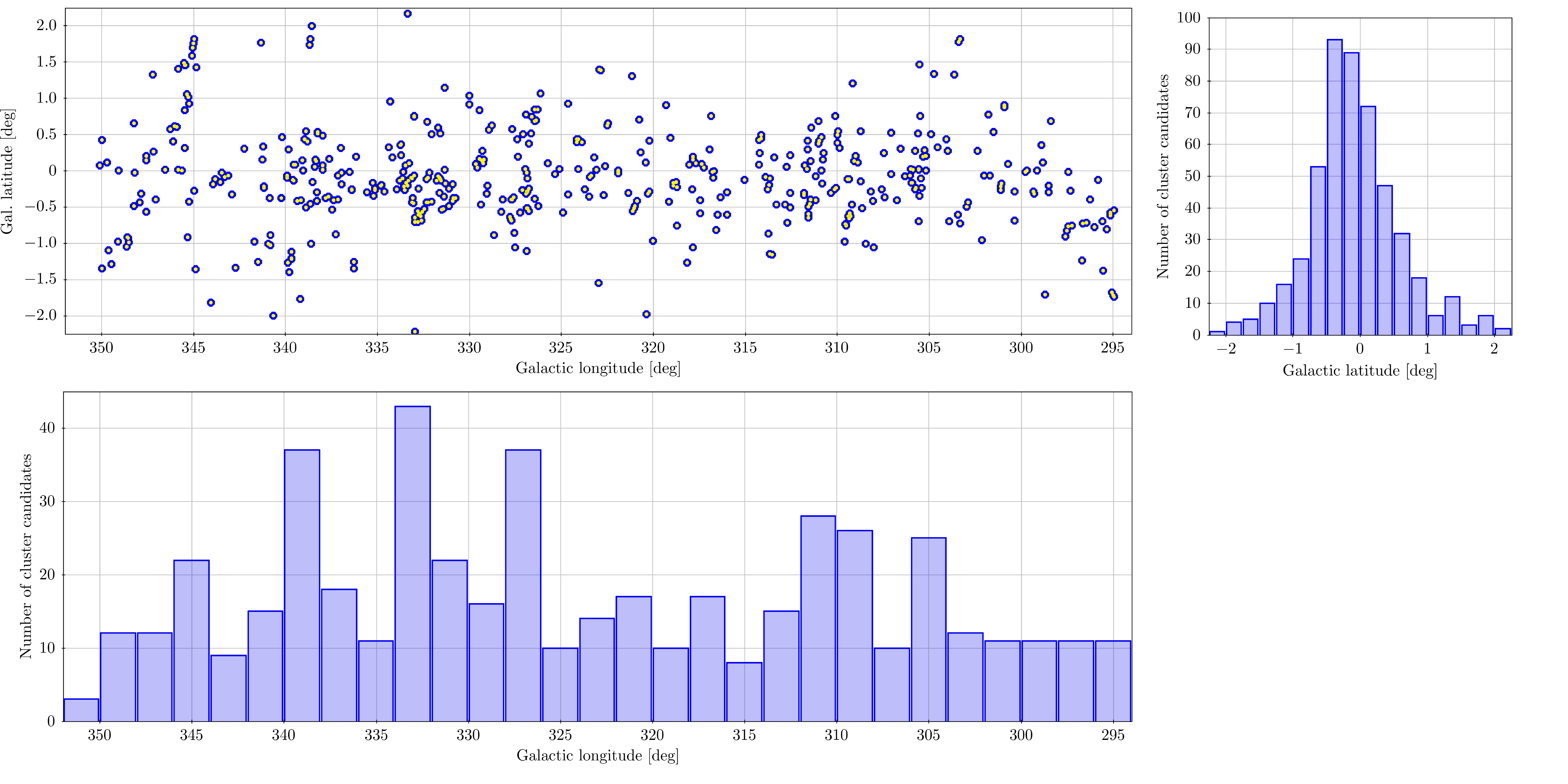}
\caption{Angular distribution of the newly discovered La Serena star
  cluster candidates in the Galactic plane (upper left panel). Also, the
  histograms of the distributions of clusters in Galactic latitude (upper
  right panel) and Galactic longitude (lower panel).}
  \label{VVV_distro}
\end{figure*}

In Table~\ref{tab1} we present a sample of the catalog, as an example of its
form and content.
The full catalog includes 493 star cluster candidates, and it will be hosted
by CDS\footnote{http://cdsweb.u-strasbg.fr/cgi-bin/qcat?J/A+A/} database and
also the VVV-ULS project page.  
Characteristics listed in columns 7 to 11 will be described in the following
sections.  
The catalog is organized as follows: 

\begin{description}
\item[{\bf Column\,1}] lists the star cluster candidate designation 
  La~Serena and running numbers from 1 to 493. 
\item[{\bf Columns\,2 to 5}] gives the equatorial (J2000.0 epoch) and galactic
  coordinates of the cluster center. These coordinates are estimated from the
  bin center, and in many cases they are corrected with inspection of the
  images by eye.
\item[{\bf Column\,6}] indicates the VVV tile where the cluster is located. 
  In a few cases, when the cluster is close to the edge of the tile, it is
  possible to find it in two different tiles.
\item[{\bf Column\,7}] lists the IRAS source associated with the cluster.
\item[{\bf Column\,8}] lists the MSX source associated with the cluster.
\item[{\bf Column\,9}] indicates the morphologic appearance of the cluster: O
  (open cluster) or C (compact cluster). Compact clusters are small
  clumps of stars, with sizes less than $30\arcsec$, mostly associated with 
  compact nebulosities in the VVV and/or GLIMPSE $8\,\mu$m images, while on
  the other hand, larger extended clusters are classified as open clusters. 
\item[{\bf Column\,10}] describes the size of the clusters based on an
  upper-limit estimation of the number of sources ($N_{\rm s}$) associated
  with the cluster. Star clusters with $N_{\rm s}<20$ are classified as small
  (S). Those with $20<N_{\rm s}<50$ are classified as medium (M). Finally,
  clusters with $N_{\rm s}>50$ are classified as large (L). We estimate
  these upper limits from the star count in the area of the cluster, doing an
  eyeball estimation of the size of the cluster. These numbers are for
  guidance because it must be noted that no background or crowding correction
  were applied.    
\item[{\bf Column\,11}] describes the morphology of each cluster based on
  $8\,\mu$m band images obtained by the Spitzer Space Telescope, under GLIMPSE
  survey. There are four main morphological types: point sources (P), knots
  (K), bubbles (B or sB), and nebula (N). Additional comments
  about the morphology of the clusters are annotated in Table~\ref{tab2}.
  Point sources refer directly to a group of point sources without
  noticeable nebulosity.
  Knot refers to a remarkable and compact nebulosity associated with 
  point sources in the $8\,\mu$m images.
  Bubble refers to arc-shaped nebulosity associated with the point
  sources. In some cases the arcs or filaments form small circles delimiting
  small cavities in the dusty clouds, while in other cases, the arcs define
  larger bubbles. For the case of small bubbles, they are marked as sB.
  Nebula refers to diffuse nebula without a specific shape associated
  with the point sources.
\end{description} 

Also, we present in Table~\ref{tab2} a sample of the association of different
astrophysical sources with each cluster candidate obtained in the
cross-matching procedures described in the Section 4.2. Again the full table
will be available in the CDS database.
Table~\ref{tab2} lists the \cite{2014A&A...562A.115S} object name (Column 2)
and the different types of objects associated with the new star cluster
candidates: dark clouds (Column 3), young stellar object candidates (Column
4), extended green objects and outflows (Column 5), and masers, with the
corresponding bibliographic references (Columns 6 and 7).

Additionally, in the Table~\ref{tab3} are included special notes and comments
for each cluster candidate. 

We have prepared a website with the detailed astrometric and morphologic
information for the 493 cluster candidates, including color-composite charts.
This catalog is available at 
{\tt http://astro.userena.cl/science/lsclusters.php} and it is open to the 
community. 
These color-composite images were created by combining $J$-band, $H$-band, and
$K_{\rm S}$-band images following the usual scheme of blue, green, and red
channels.  
Each color image was cropped and dynamically stretched to display each cluster
comfortably. 
For example, the chart for the compact cluster La~Serena 309 was cropped
to a size of $1\arcmin \times 1\arcmin$, while the chart corresponding to the
open cluster candidate La~Serena 109 was cropped to a 
$3\arcmin \times 3\arcmin$ field of view, respectively.   
As example, Figure~\ref{vvv-laserena_examples} presents $JHK_{\rm S}$
color-composite images of some of the newly discovered star cluster
candidates.  
The complete set of color-composite chart images will be available in 
VVV-ULS web page.    

\section{Characteristics of the new cluster candidates}

\subsection{Spatial distribution}

The angular distribution of the new stellar group candidates
(Figure~\ref{VVV_distro}, upper left panel) shows an obvious concentration 
toward the Galactic plane. 
The distribution in Galactic latitude is characterized by a single-peaked
distribution with a half-width at half-maximun of about $0.6^\circ$
(Figure~\ref{VVV_distro}, upper right panel). 
The peak of distribution is slightly shifted $-0.2^\circ$ from the Galactic
 plane.  
The distribution in Galactic longitude is noticeable based on the presence of
 many groups of clusters associated with large star-forming
complexes (Figure~\ref{VVV_distro}, lower panel).   
For example, clusters located at $l=305\degr-306\degr$ are associated to
the star-forming complex \object{G305}; at $l= 309\degr$ are associated with the
\ion{H}{ii} region \object{RCW\,80}, clusters located at $l=310\degr$ with the
\ion{H}{ii} region \object{RCW\,83}, at $l=326\degr$ with the star-forming
complex \object{RCW\,95} and \object{RCW~97}, at $l=333\degr-334\degr$ with
the star-forming complex \object{RCW\,106}, 
at $l=339\degr$ with \ion{H}{ii} region \object{RCW\,109}, and the small
maximum of cluster count located at $l=346\degr$ with the giant molecular
complex \object{GMC 345.5+1.0} \citep[][]{2011A&A...534A.131L}.   
We notice a decrease in the number of discovered clusters at $l>340\degr$,
which may be related to an increase in extinction and/or increased crowding
toward the inner disk, which dilutes the ability to detect poor star clusters. 

\subsection{Spatial correlation with other infrared sources and  
                 structures}  

The methodology used to detect candidates certainly favors the discovery of
compact groups.  
Detected densities were boosted in regions near or in the dark clouds, where
the number of field stars drops owing to the presence of these clouds. 
Most of these cluster candidates are compact groups whose stars undergo high
reddening, and many times they have associated small emission nebulosities. 
Some examples of this kind of clusters are La~Serena 045, La~Serena 185, and
La~Serena 328. 
At first glance, the morphology of these star cluster candidates suggests that
they are in the stage of ongoing star formation.  
Therefore, if the star formation processes are in such early stages, then many
of these clusters could be associated with mid-infrared sources. 

Therefore we performed a number of cross-matching procedures between our
catalog and the different catalogs available in SIMBAD and Vizier in order  
to identify the mid-infrared (mid-IR) source counterparts. 
The cross-matching procedures utilize TopCat virtual observatory tool
\citep{2005ASPC..347...29T}.

\begin{figure}[h!]
   \centering
\includegraphics[width=8cm]{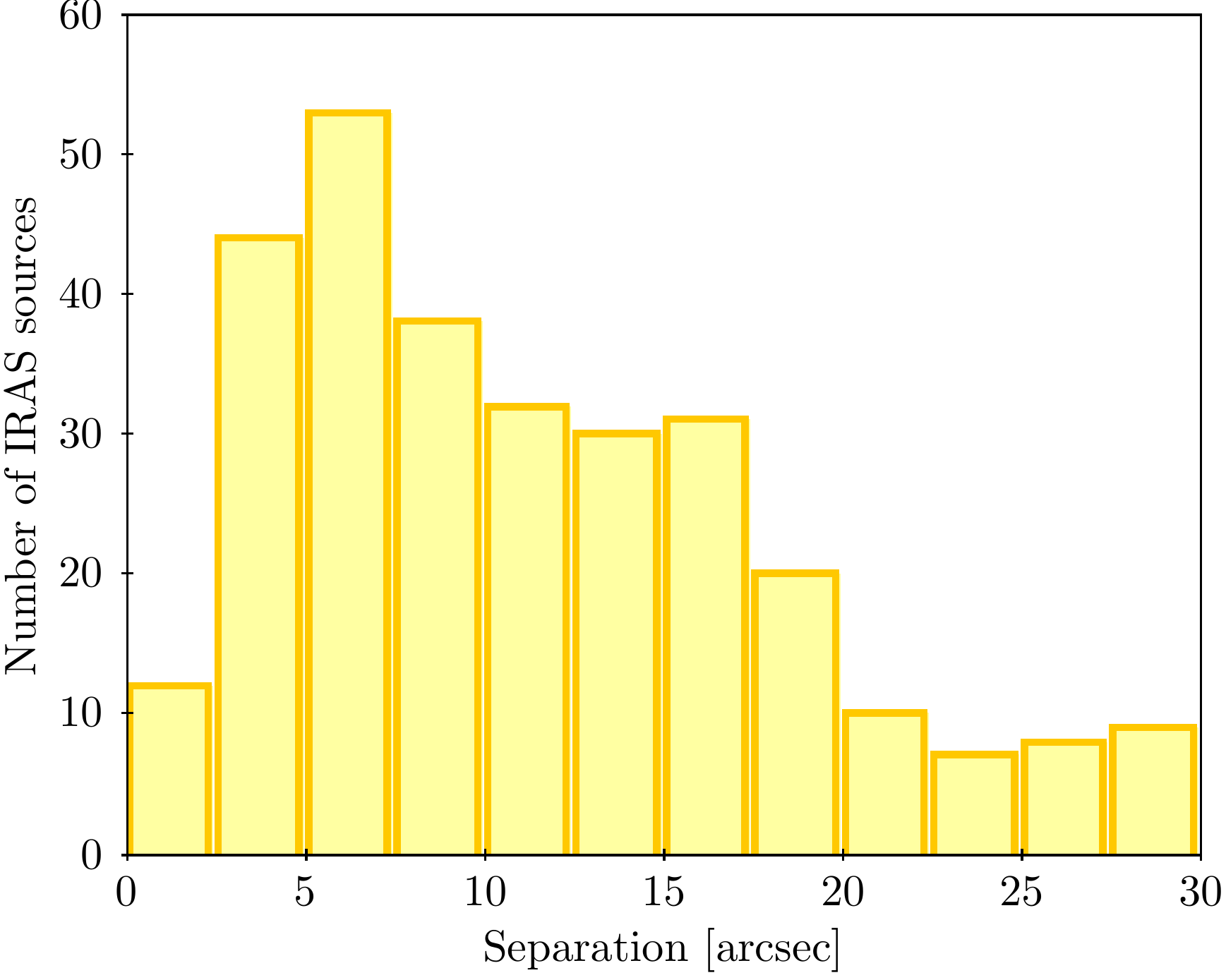}
  \caption{Histogram of the number of IRAS point sources associated with VVV--La
    Serena star cluster candidates.}
   \label{xmatch-histogram-iras}
\end{figure}

\begin{figure}[h!]
   \centering
\includegraphics[width=8cm]{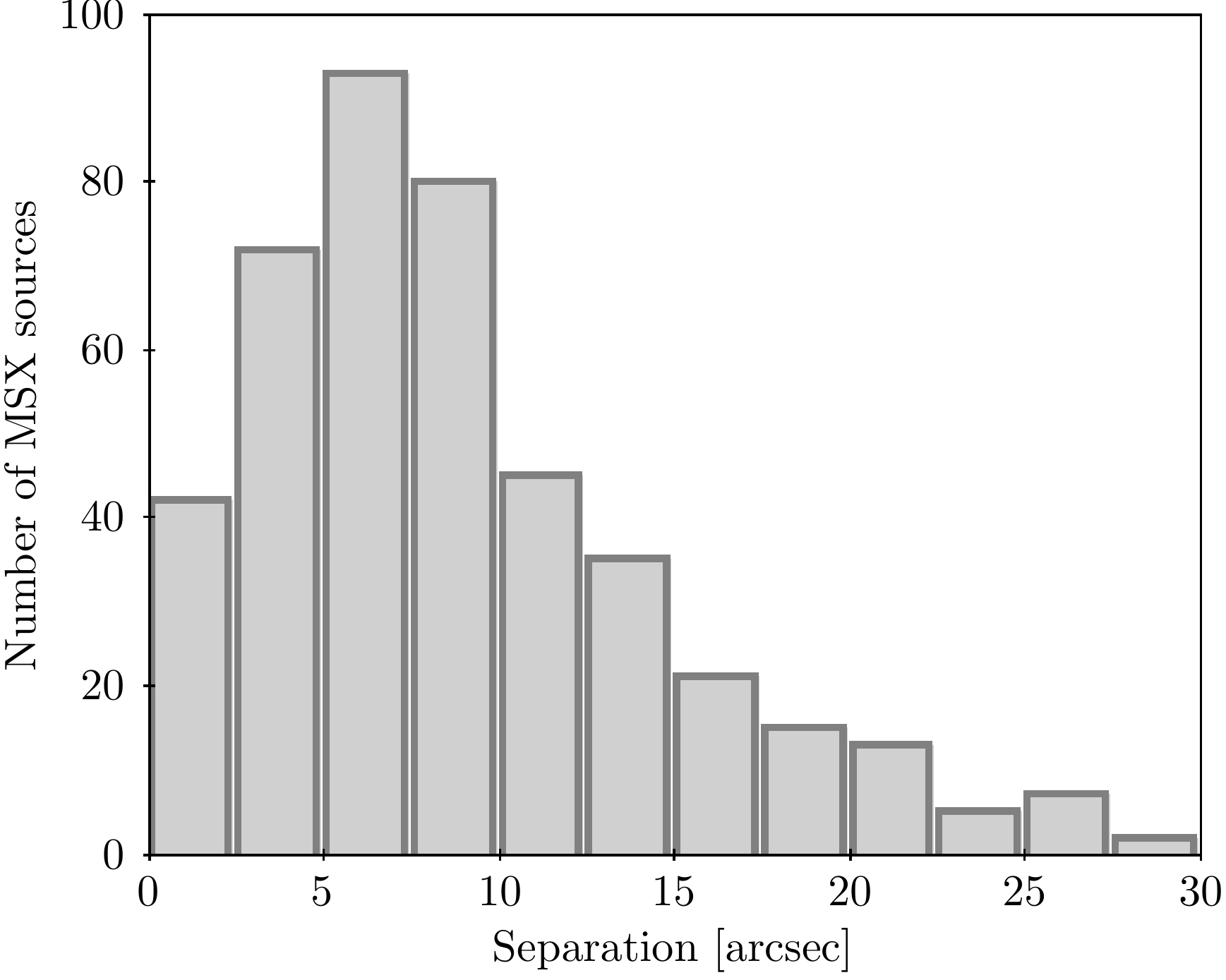}
  \caption{Histogram of the number of MSX sources associated with VVV -- La
    Serena star cluster candidates.}
   \label{xmatch-msx-clusters}
\end{figure}

The first two cross-matching procedures are based on the Infrared Astronomical
Satellite (IRAS) point source catalog (PSC) \citep{1988iras....1.....B} and
the {\em Midcourse Space Experiment} (MSX) catalog version 6 (MSX6C)
\citep{2003yCat.5114....0E}. 
IRAS PSC includes about 250,000 reliable infrared point sources observed by
the Infrared Astronomical Satellite. 
These sources have angular extents less than approximately 0.5, 0.5, 1.0, and
2.0 arcmin in the in-scan direction at 12, 25, 60, and $100\,\mu$m,
respectively. 
Away from confused regions of the sky, the survey is complete to about 0.4,
0.5, 0.6, and 1.0 Jy at these wavelengths. 
The representative  position uncertainties are about $2\arcsec$ to $6\arcsec$
in-scan and about $8\arcsec$ to $16\arcsec$ cross-scan directions,
respectively. 
The MSX6C in the Galactic plane includes about 430,000 point sources observed
with the infrared instrument SPIRIT III on MSX, with an astrometric accuracy
better than $1\arcsec$.  
This instrument consists of a 33 cm clear-aperture off-axis telescope with
five line scanned infrared focal plane arrays of $18.3\arcsec$ square pixels,
and a high sensitivity (0.1\,Jy at $8.3\,\mu$m).  
SPIRIT III is provided with six photometric bands: B1 ($4.29\,\mu$m), B2
($4.35\,\mu$m), A ($8.28\,\mu$m), C ($12.13\,\mu$m), D ($14.65\,\mu$m), 
and E ($21.34\,\mu$m).  

We find 293 IRAS sources closer than a cross-matching radius of $30\arcsec$
(Figure~\ref{xmatch-histogram-iras}), corresponding to 59\% of VVV cluster
candidates.  
The matching radius used corresponds to the IRAS spatial resolution at
$12\,\mu$m; however, this may be too restrictive with respect to the IRAS
resolution, but will reduce the confusion with neighboring Mid-IR structures.
The IRAS point sources associated with each star cluster candidate are listed
in column 7 of Table~\ref{tab1}.

We also matched the cluster candidates with Midcourse Space Experiment (MSX)
catalog version 6 \citep{2003yCat.5114....0E}.  
Compared with IRAS, the MSX survey has better spatial resolution ($18\arcsec$)
and sensitivity on the Galactic plane, and  we might expect a tighter
cross-matching. 
Figure~\ref{xmatch-msx-clusters} shows the distribution of separations for the
MSX sources, using a matching radius of $30\arcsec$. 
We have 436 MSX point sources ($88\%$) associated with our star cluster
candidates, with 50\% of them inside the $8\arcsec$ circle. 
The MSX sources associated with our source catalog are indicated in column 8
of Table~\ref{tab1}.
For sparse open clusters candidates, composed presumably of red giant or
supergiant stars distributed over several square arcminutes (e.g., La~Serena
314), the MSX point source listed in Table~\ref{tab1} corresponds to the
closer source to the position derived for the cluster center.  

\begin{figure}[h!]
   \centering
\includegraphics[width=9.3cm]{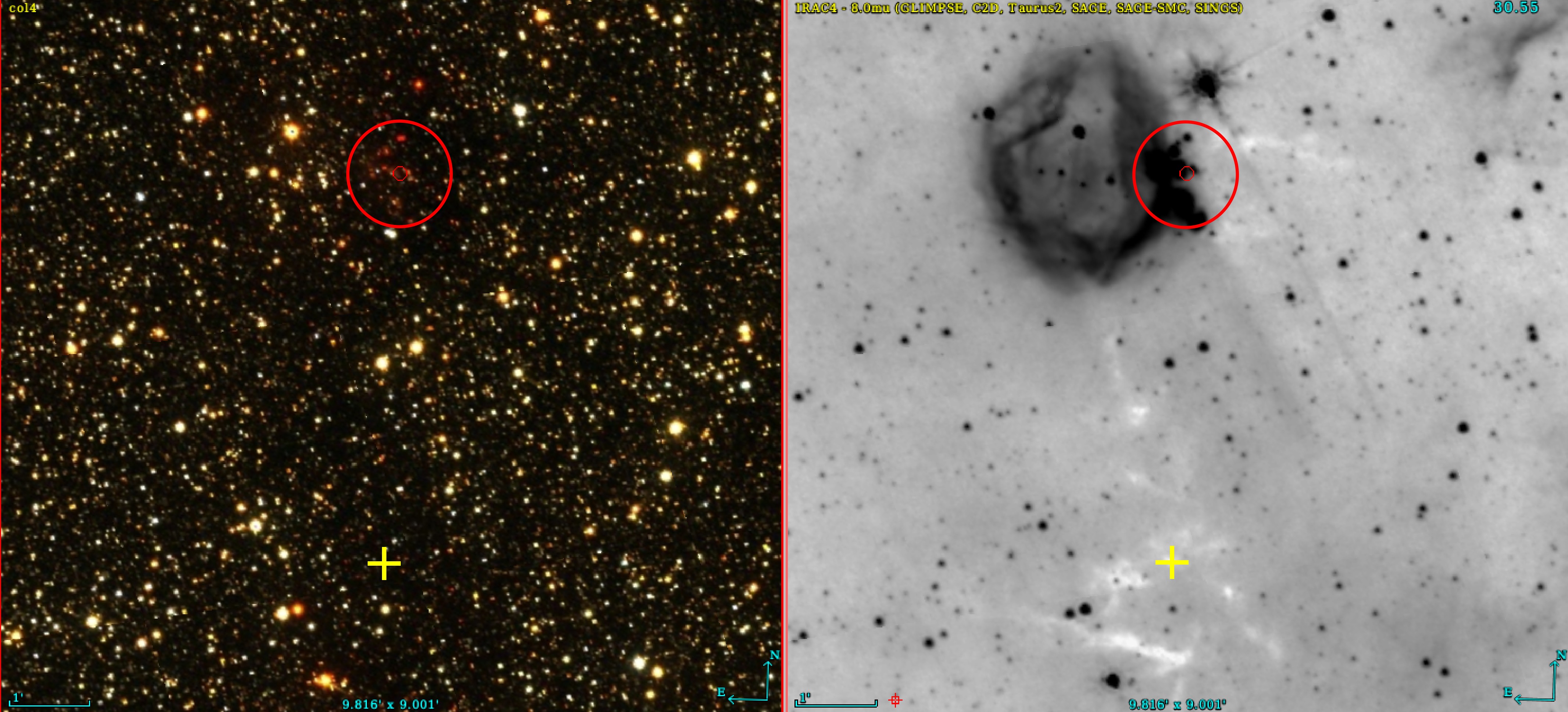}
  \caption{Area of cluster candidate La Serena 208.
    Left panel: VVV color-composite image following the same scheme as in
    Figure\,\ref{vvv-laserena_examples}. Right panel: Spitzer/GLIMPSE
    8\,$\mu$m. The cluster candidate is marked with a small red circle, and
    the dark cloud \object{SDC\,G$321.336-0.373$} wth a small yellow plus
    sign. The field of view of each panel is about $9.8 \times 9.0$ arcmin.} 
    \label{dark-nebula-ls-208} 
\end{figure}

As mentioned above, the vast majority of cluster candidates were detected in
areas with dark and/or bright nebulosities on the VVV images.  
A visual inspection of VVV images finds a total of 456 cluster candidates
associated with dark clouds up to a distance of $2\arcmin$.
We searched for dark cloud catalogs in the Vizier database in the area
covered by the Galactic disk VVV tiles, and we find four main entries:
\cite{2009A&A...505..405P}, \cite{2011PASJ...63S...1D},
\cite{2002A&A...383..631D}, and \cite{1986A&AS...63...27H}.
The first one is the most important for our study by far. 
The cross-matching between our catalog and these dark cloud catalogs brings
310 cluster candidates ($\%63$) associated with 552 dark clouds. Column 3 in
Table 2 lists the dark clouds associated with each cluster candidate in
detail.    
Interestingly, most of the dark nebulae change their appearance and morphology
with the wavelength range of the observations. 
As an example, we can describe the area surrounding the cluster La~Serena 208. 
Figure\,\ref{dark-nebula-ls-208} shows the VVV color-composite image and
Spitzer/GLIMPSE 8\,$\mu$m around this cluster. 
In the VVV color-composite image, it is possible to discern a filamentary dark
nebula that causes a marked reddening and a decline in the star count. 
The Spitzer/GLIMPSE 8\,$\mu$m image at the position of La~Serena 208 reveals a
cluster of knot point sources associated with filamentary arcs 
that are deployed eastward, forming a small bubble of $1\farcm3$ diameter,
cataloged as \object{[CPA2006] S89} \citep[][]{2006ApJ...649..759C}. 
Additionally, an infrared dark cloud connected with the bubble is located
$4\farcm8$ to the south.  
This infrared dark cloud (IRDC), known as \object{SDC\,G321.336-0.373}
\citep{2009A&A...505..405P}, has an angular size of
$0\farcm525\times0\farcm222$. 
Examination of the GLIMPSE $8\mu$m image suggests a connection between the
IRDC and the bubble through a series of uncataloged dark filaments. 

\begin{figure*}
   \centering
\includegraphics[width=45mm]{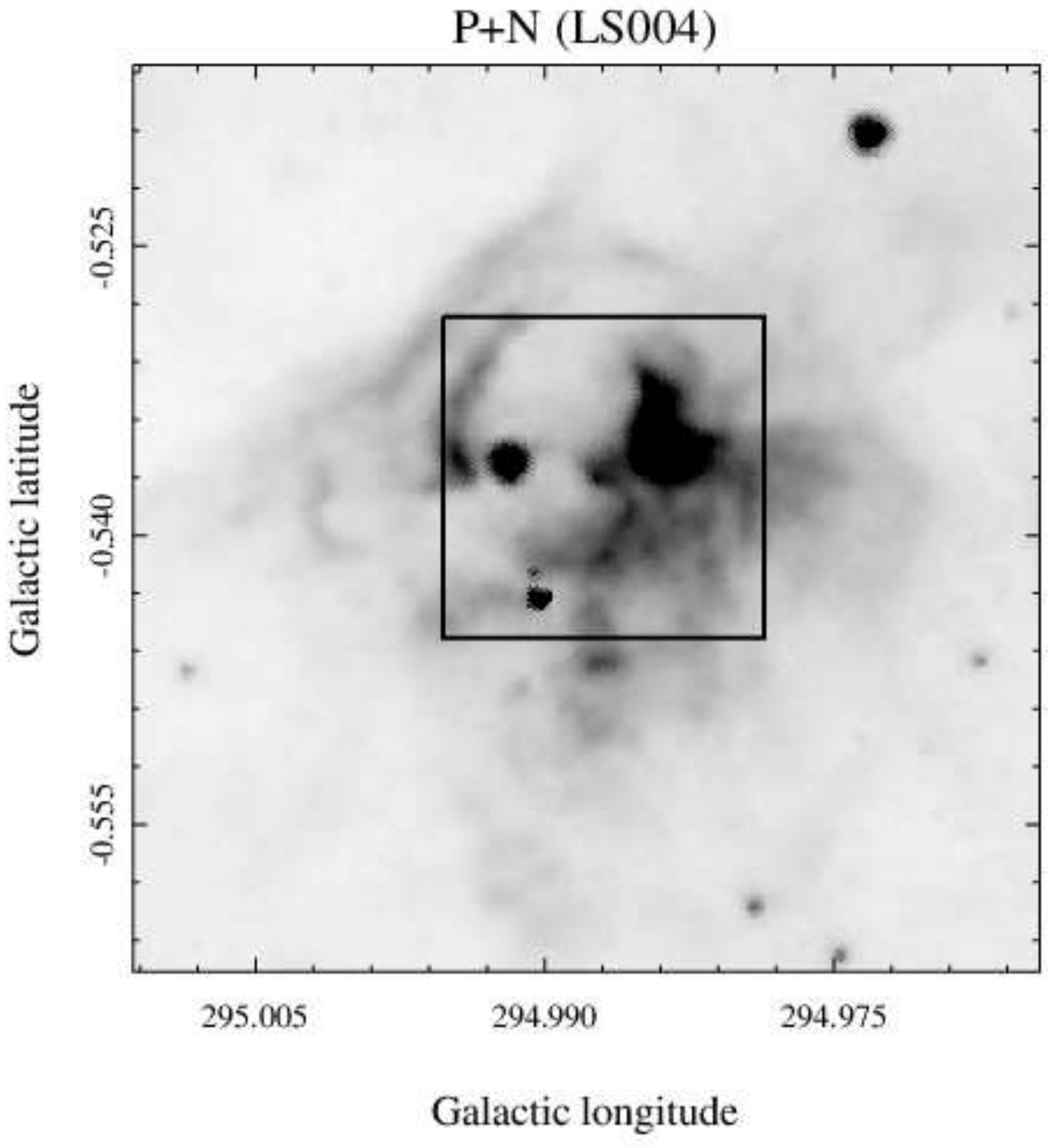}
\includegraphics[width=45mm]{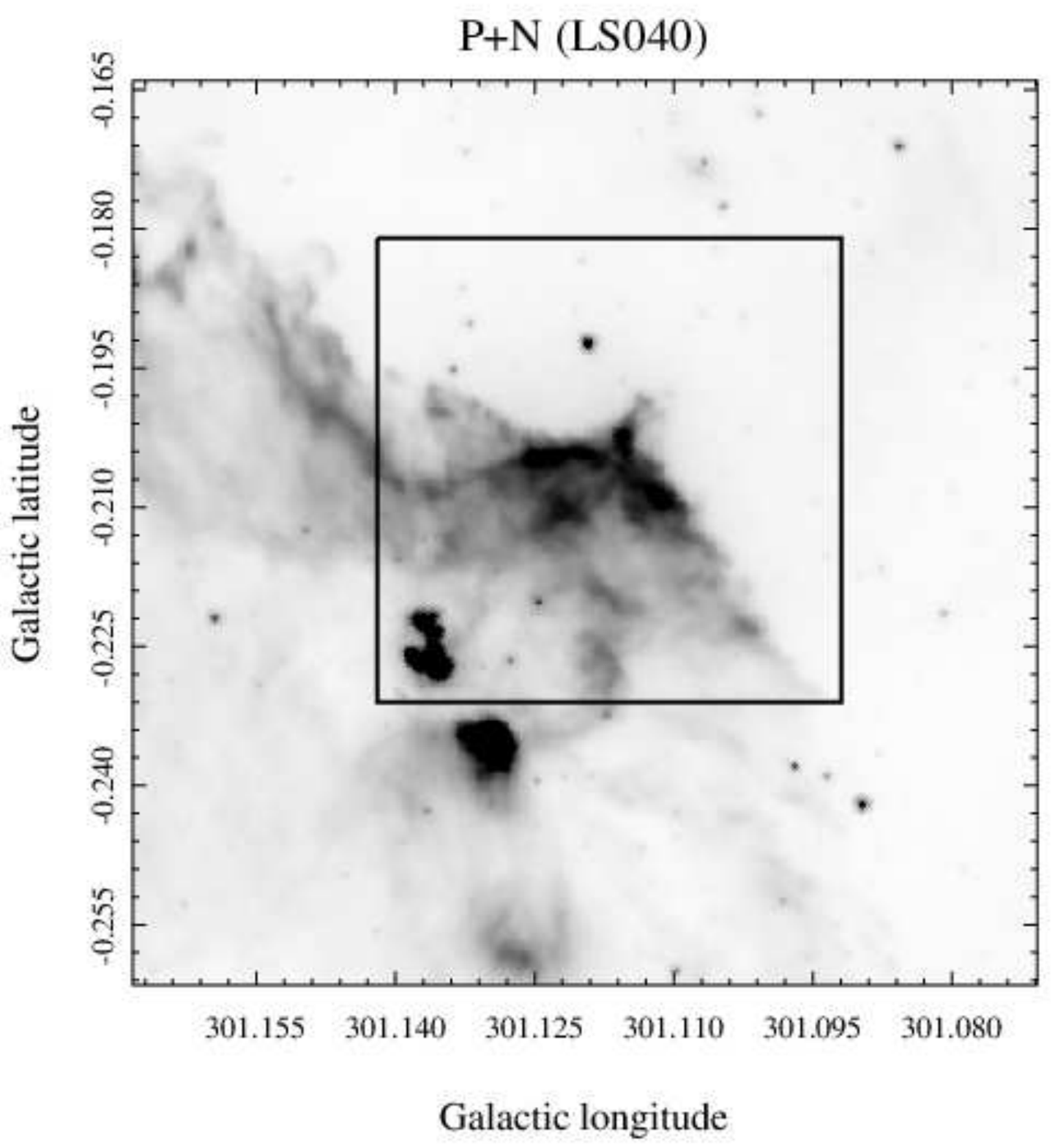}
\includegraphics[width=45mm]{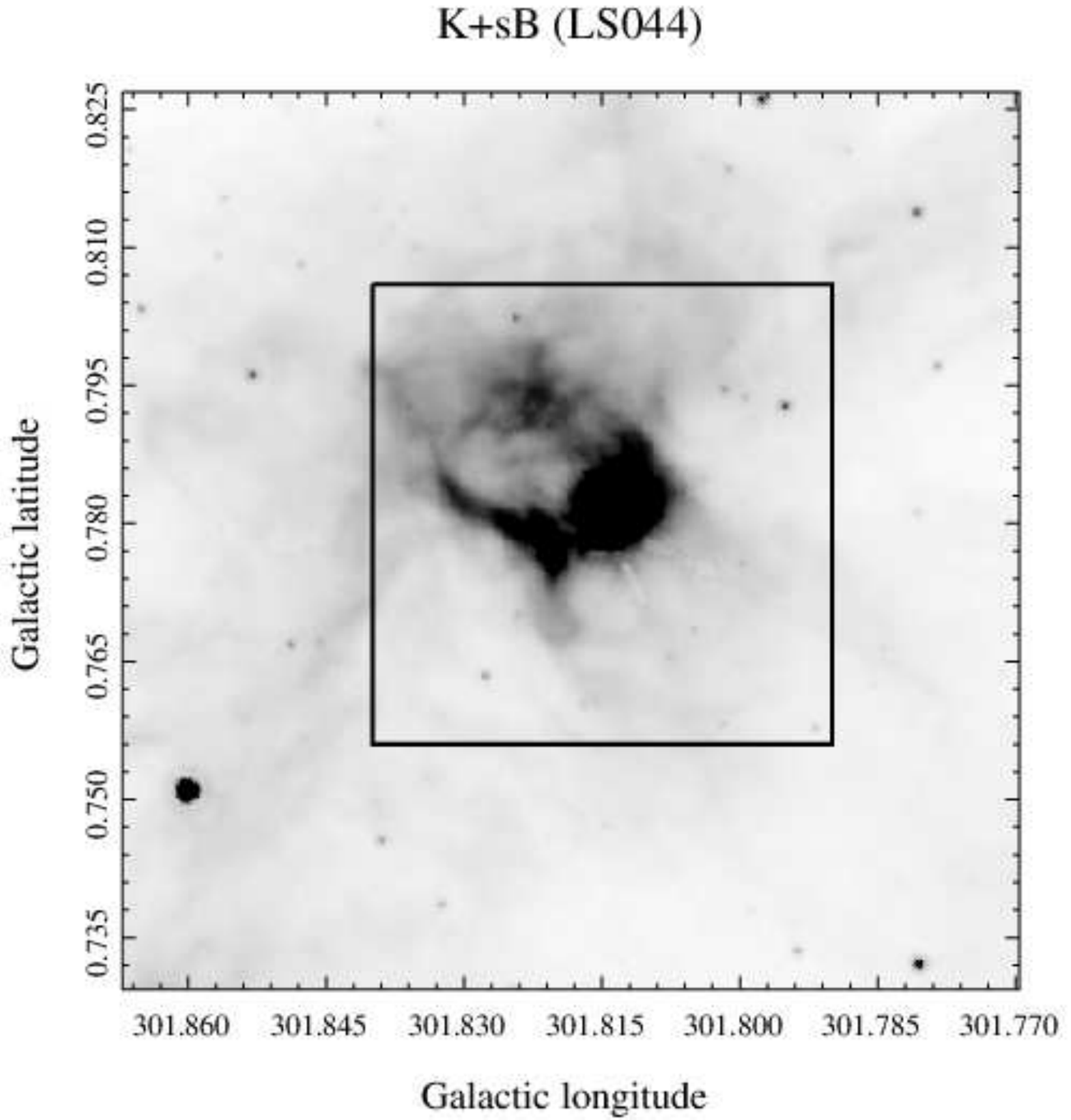}
\includegraphics[width=45mm]{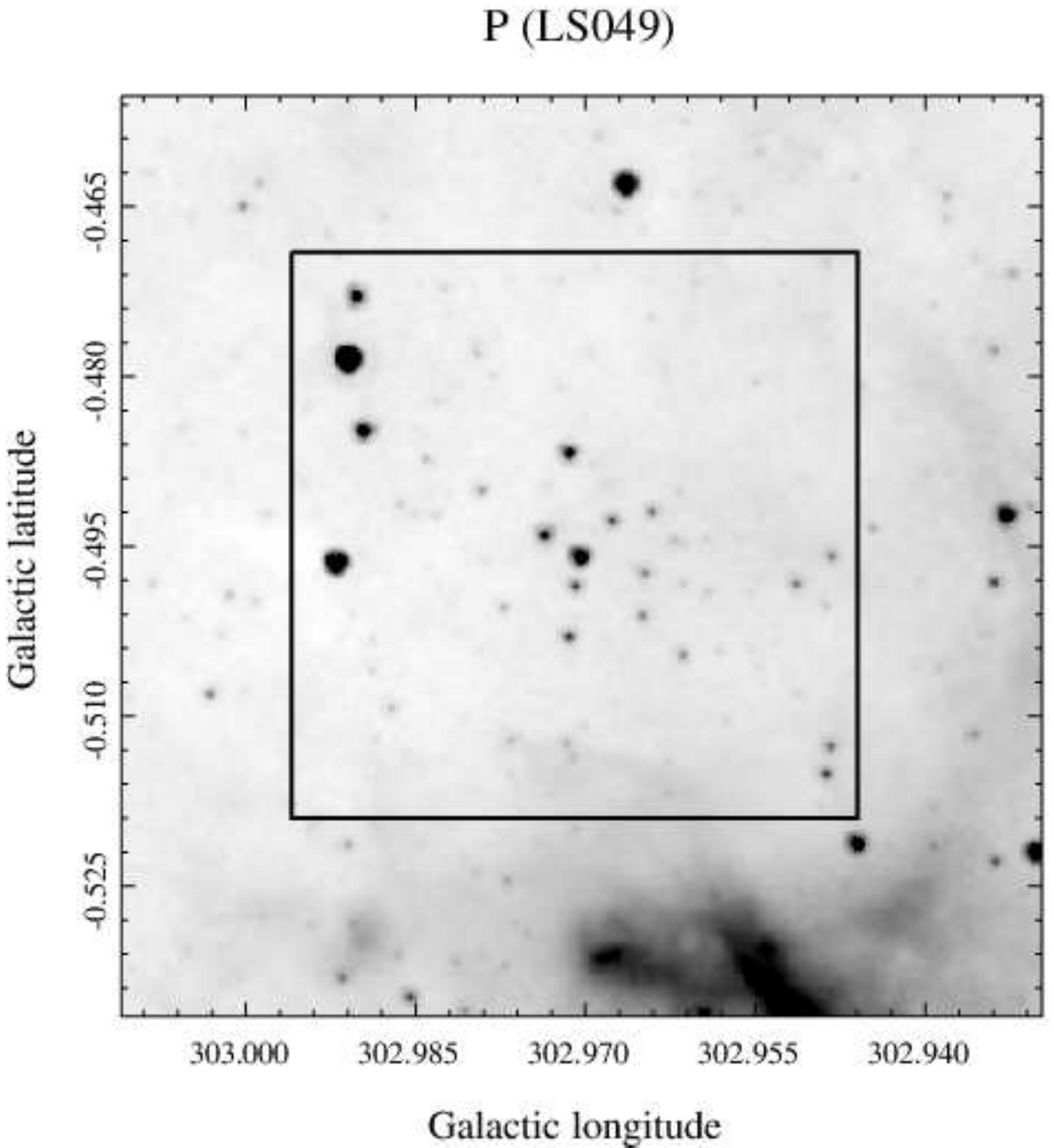}
\includegraphics[width=45mm]{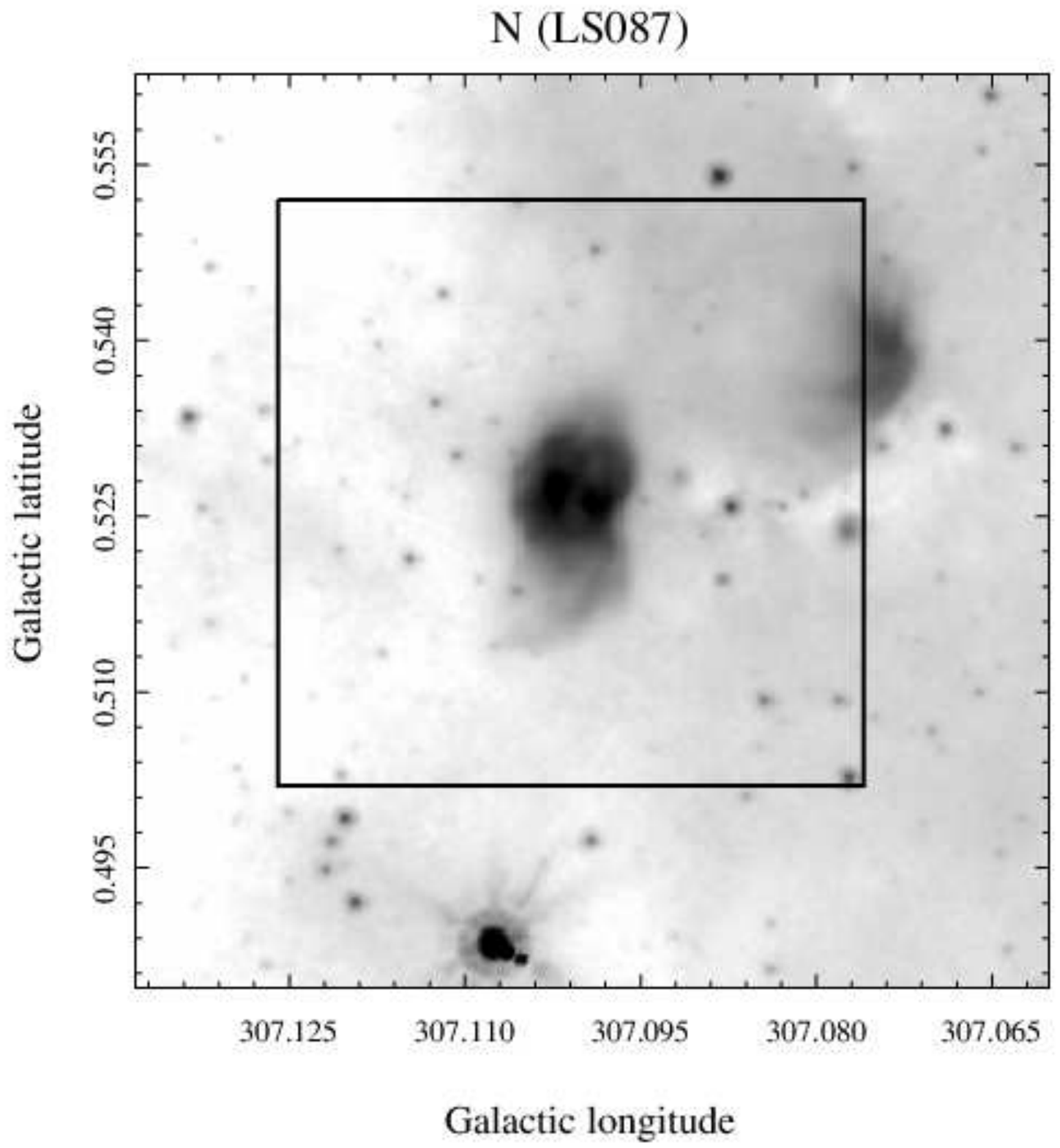}
\includegraphics[width=45mm]{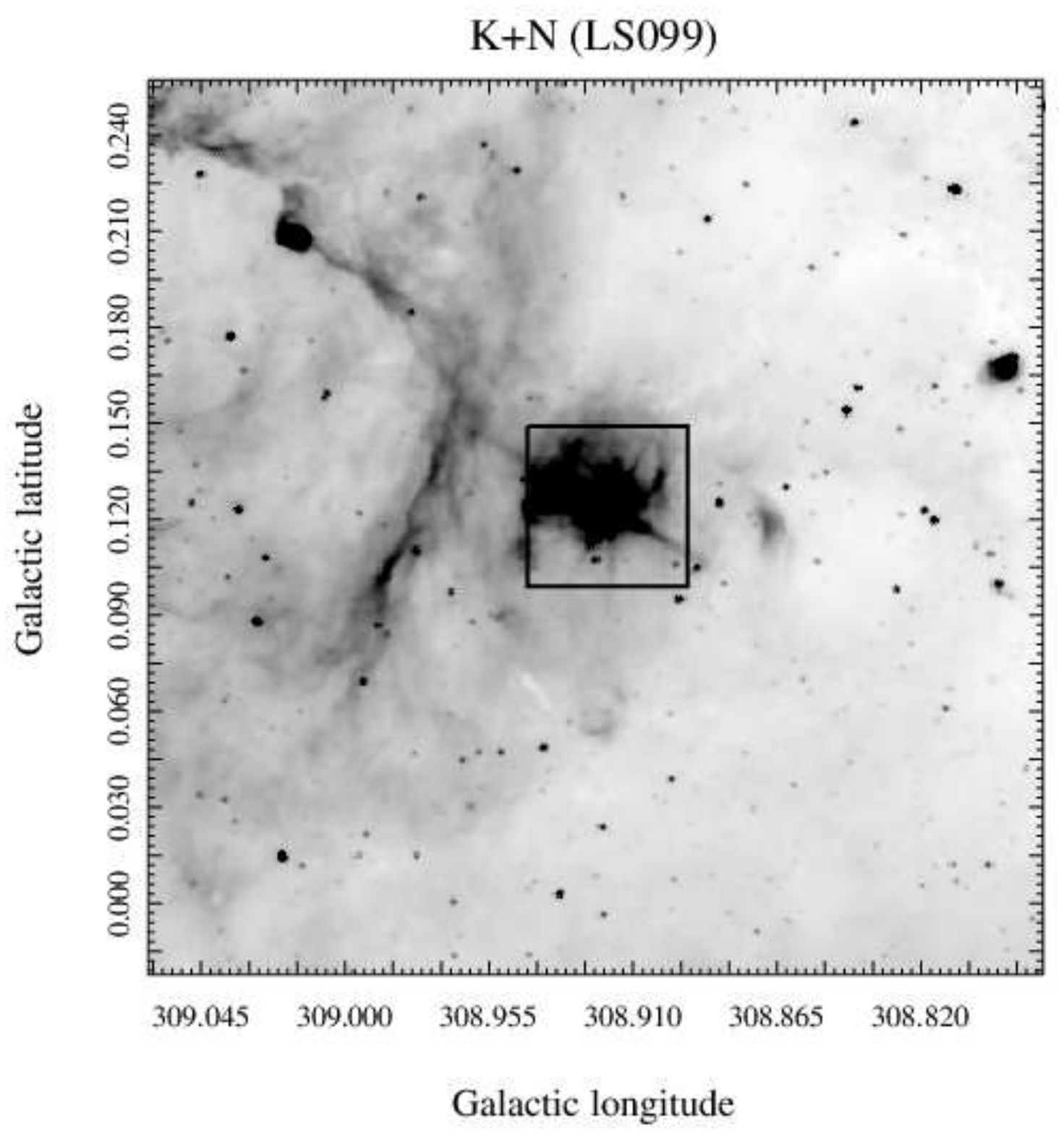}
\includegraphics[width=45mm]{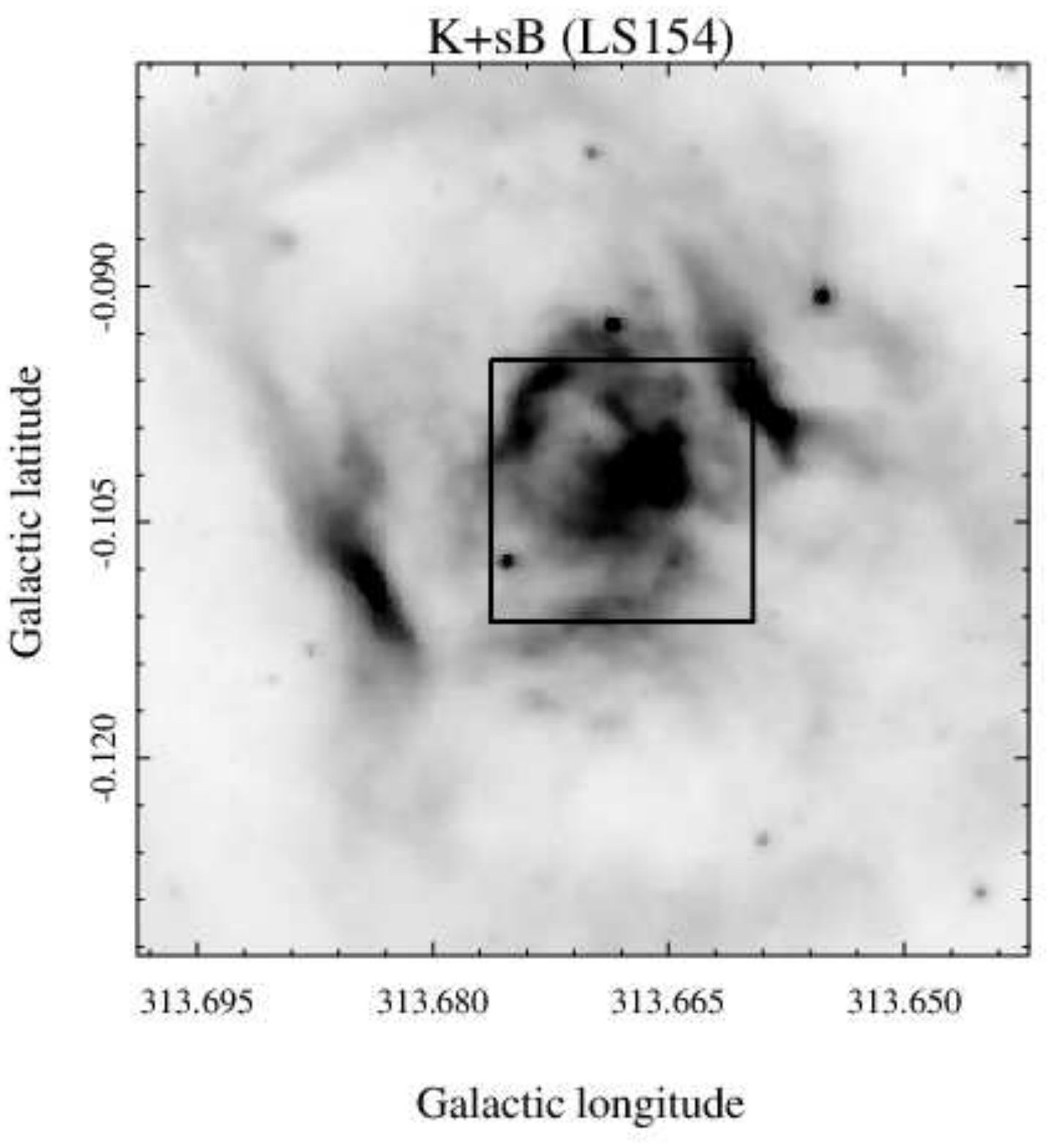}
\includegraphics[width=45mm]{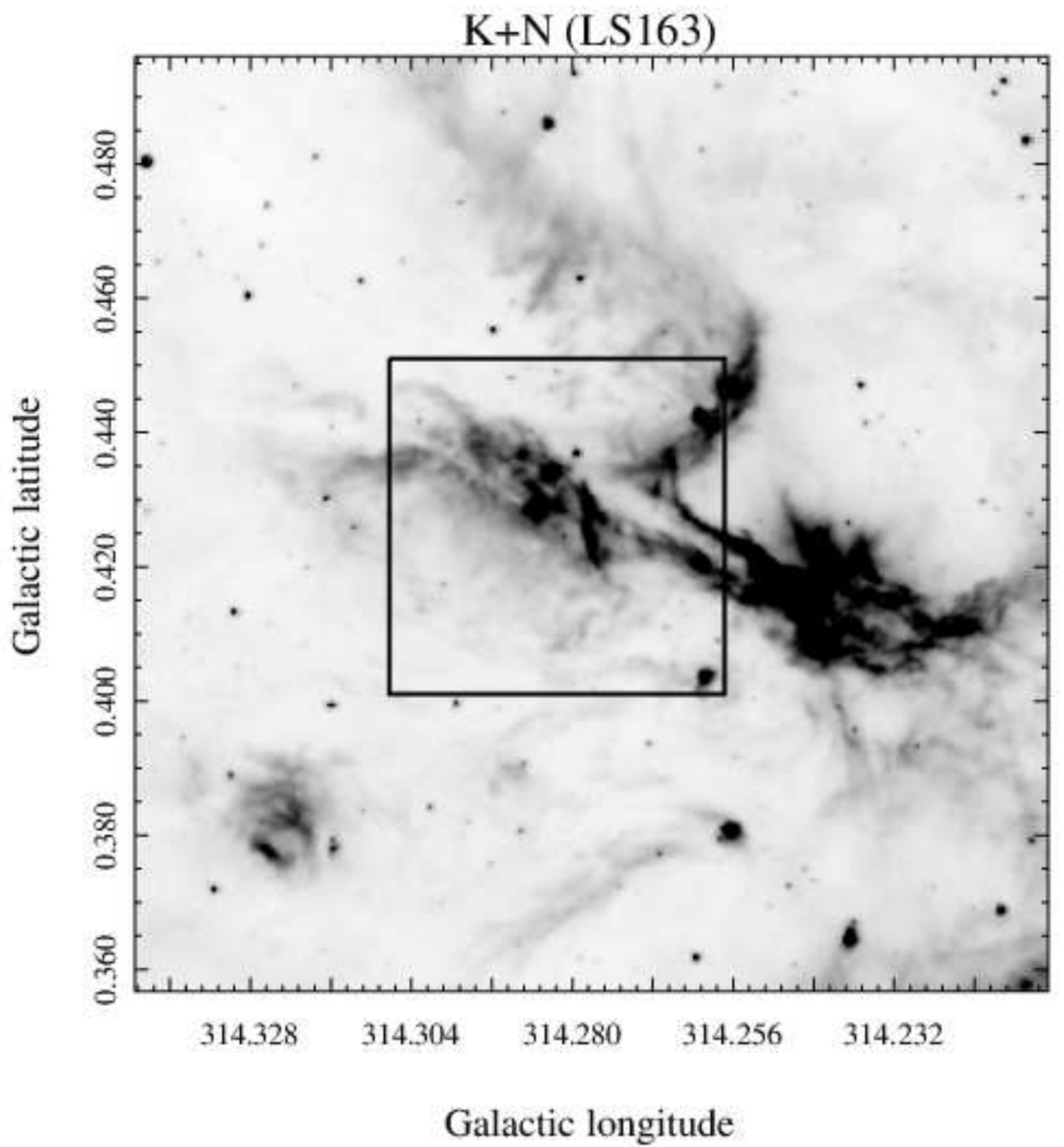}
\includegraphics[width=45mm]{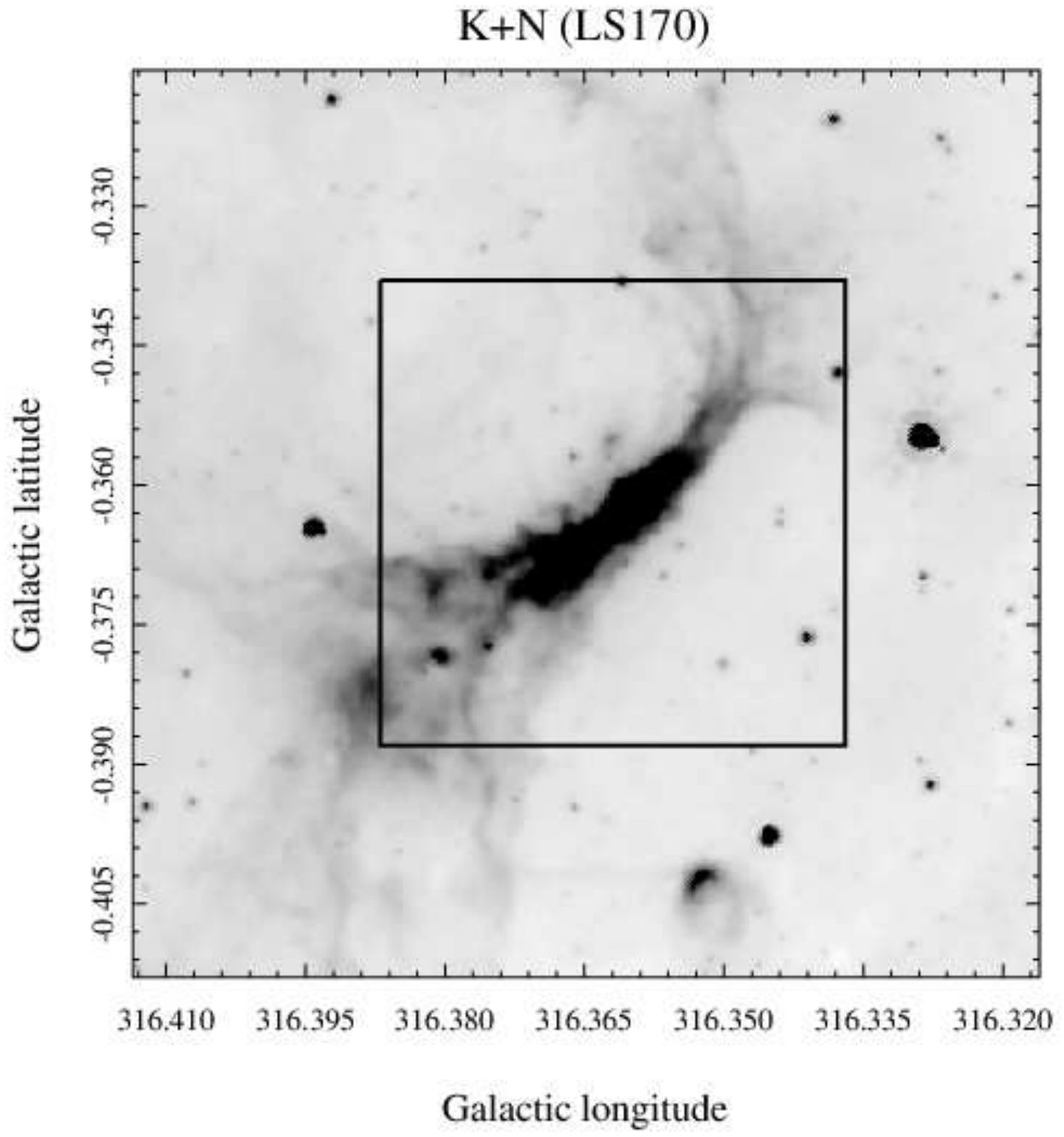}
\includegraphics[width=45mm]{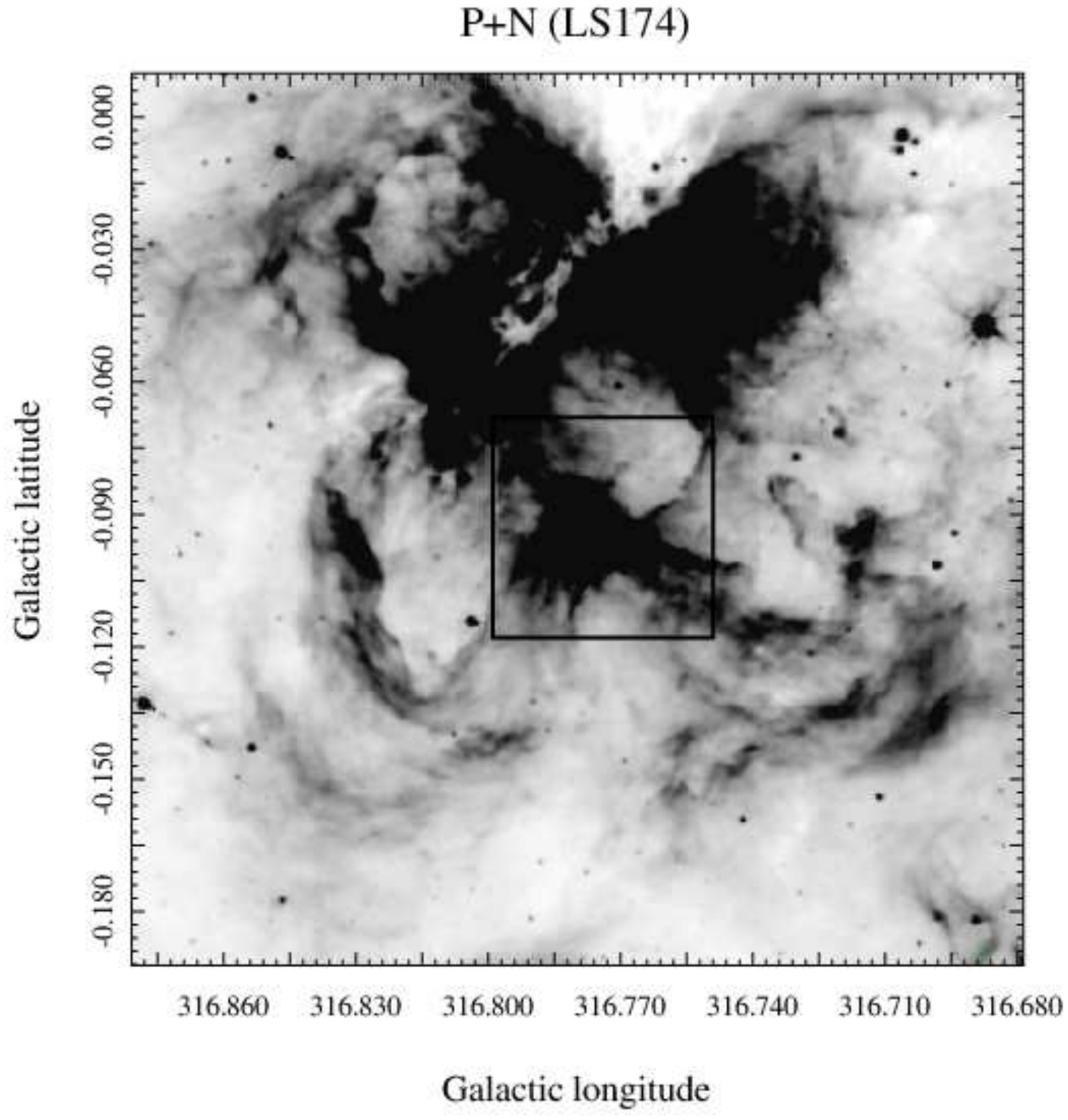}
\includegraphics[width=45mm]{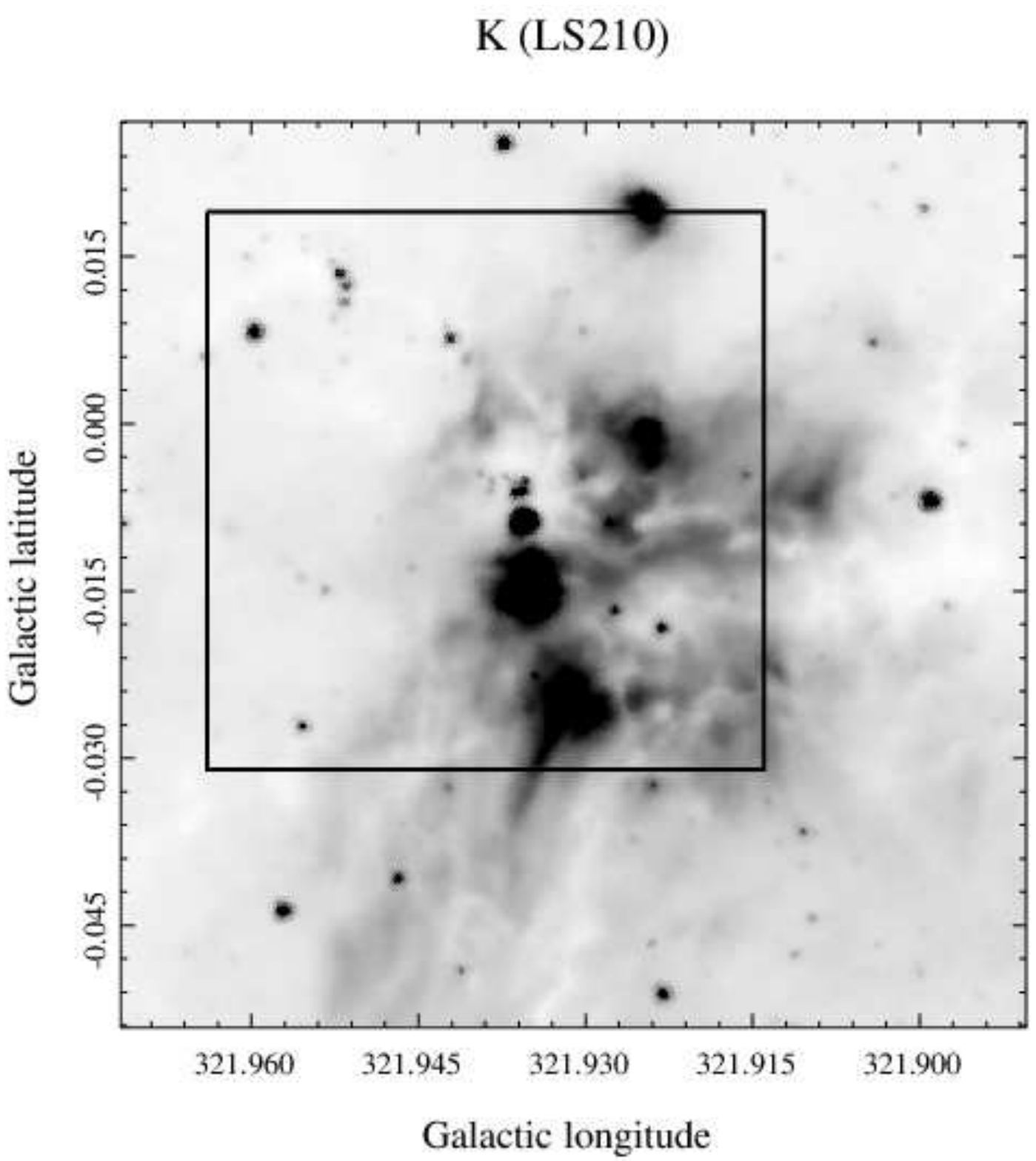}
\includegraphics[width=45mm]{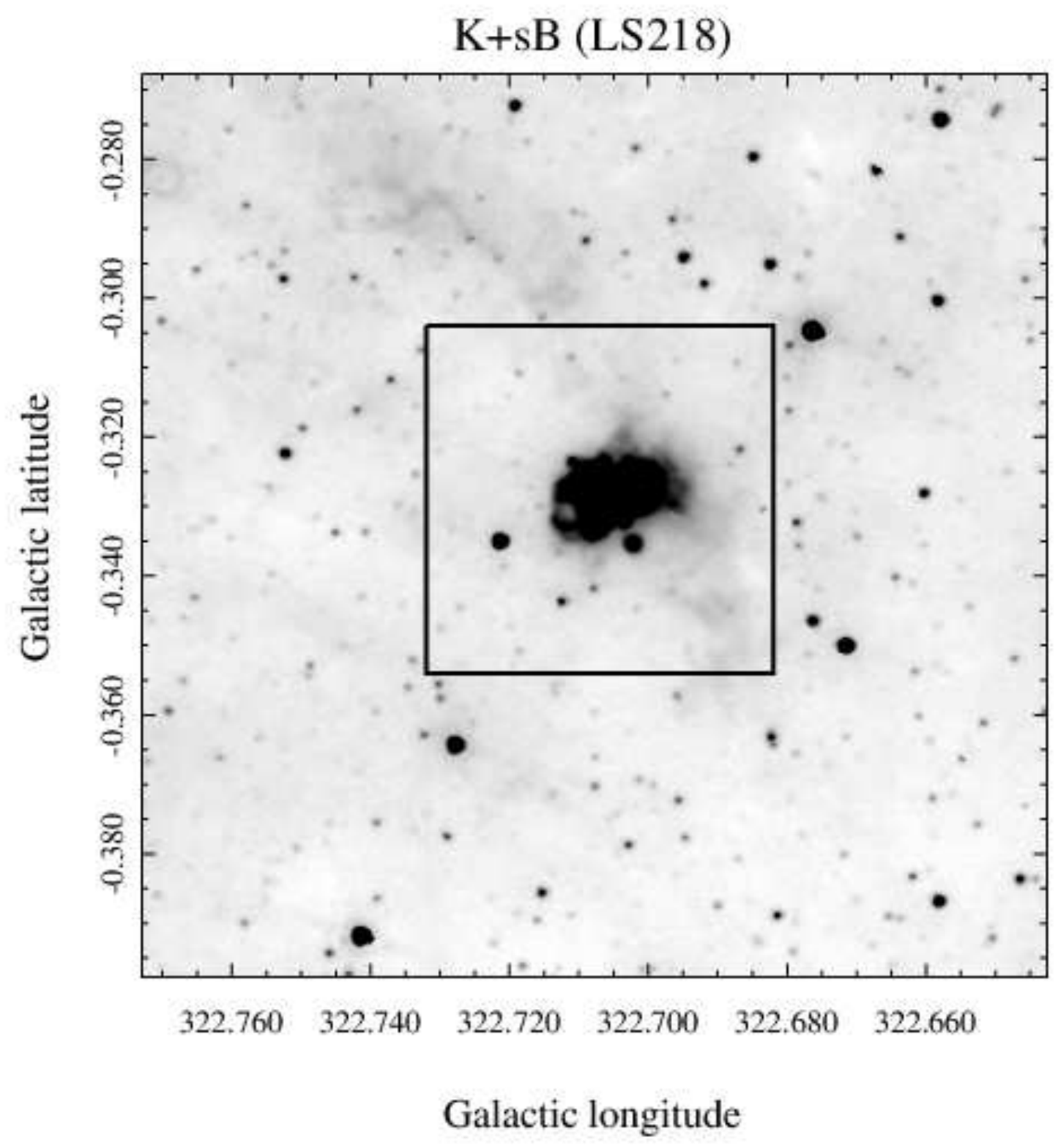}
\includegraphics[width=45mm]{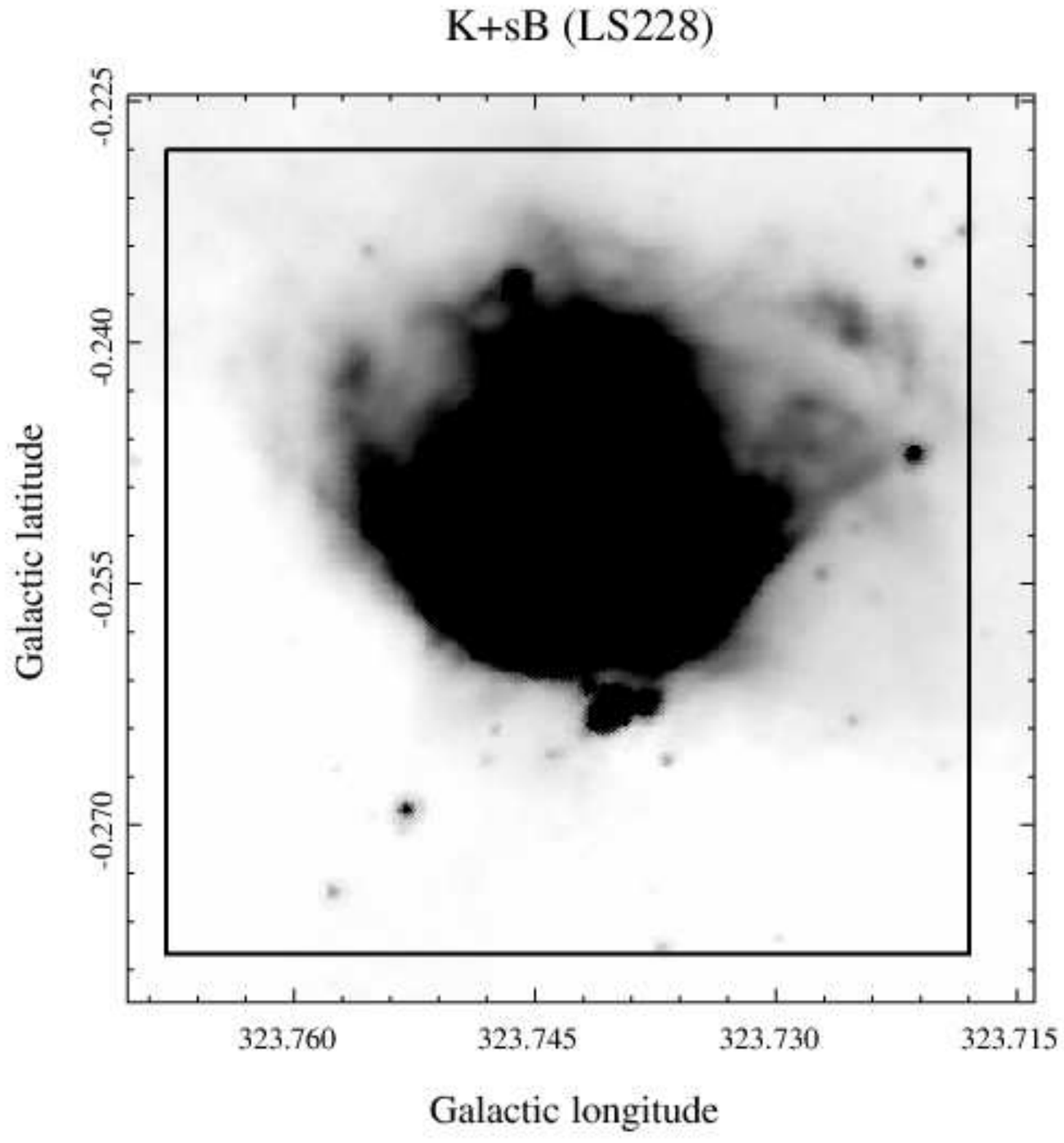}
\includegraphics[width=45mm]{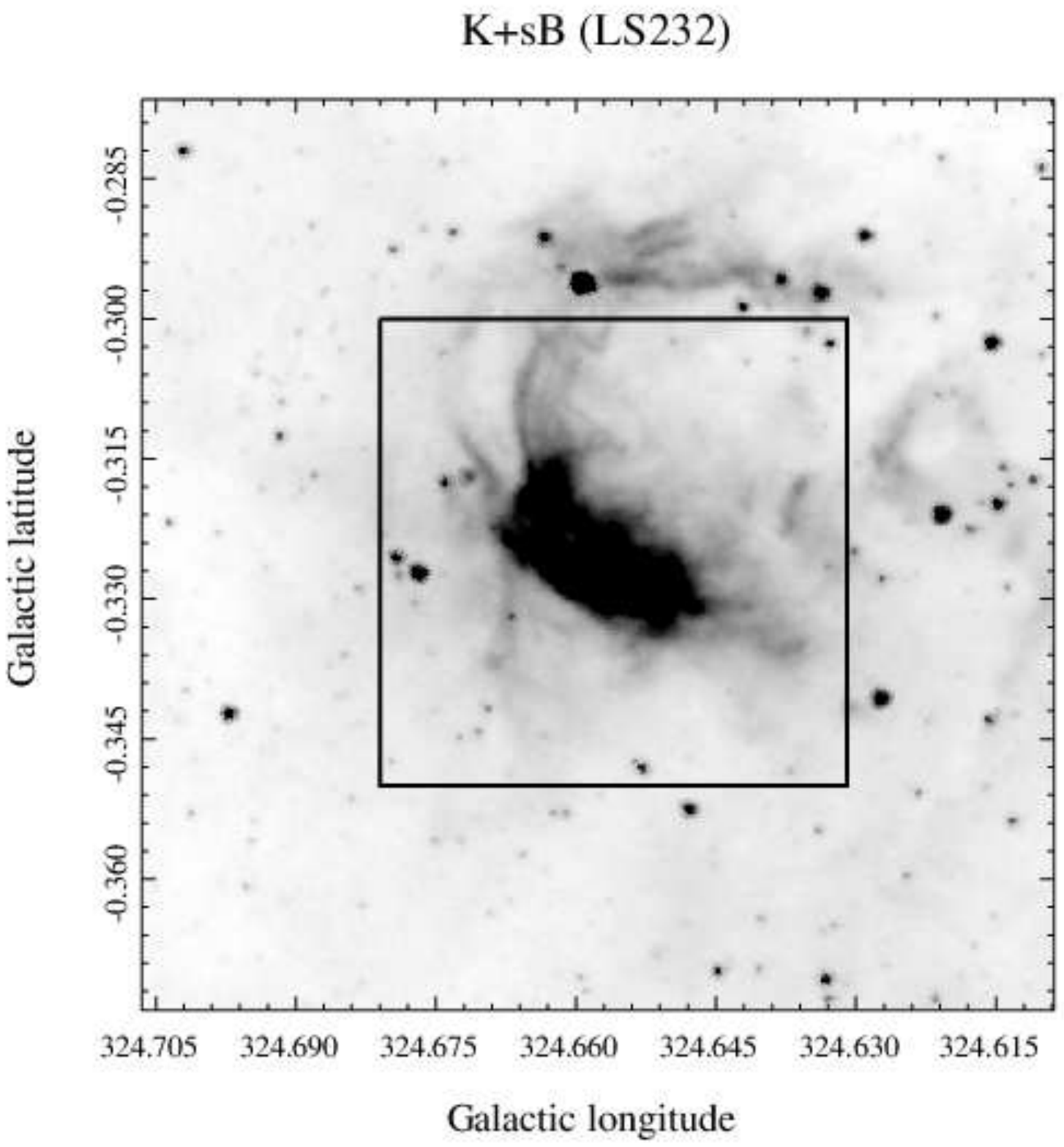}
\includegraphics[width=45mm]{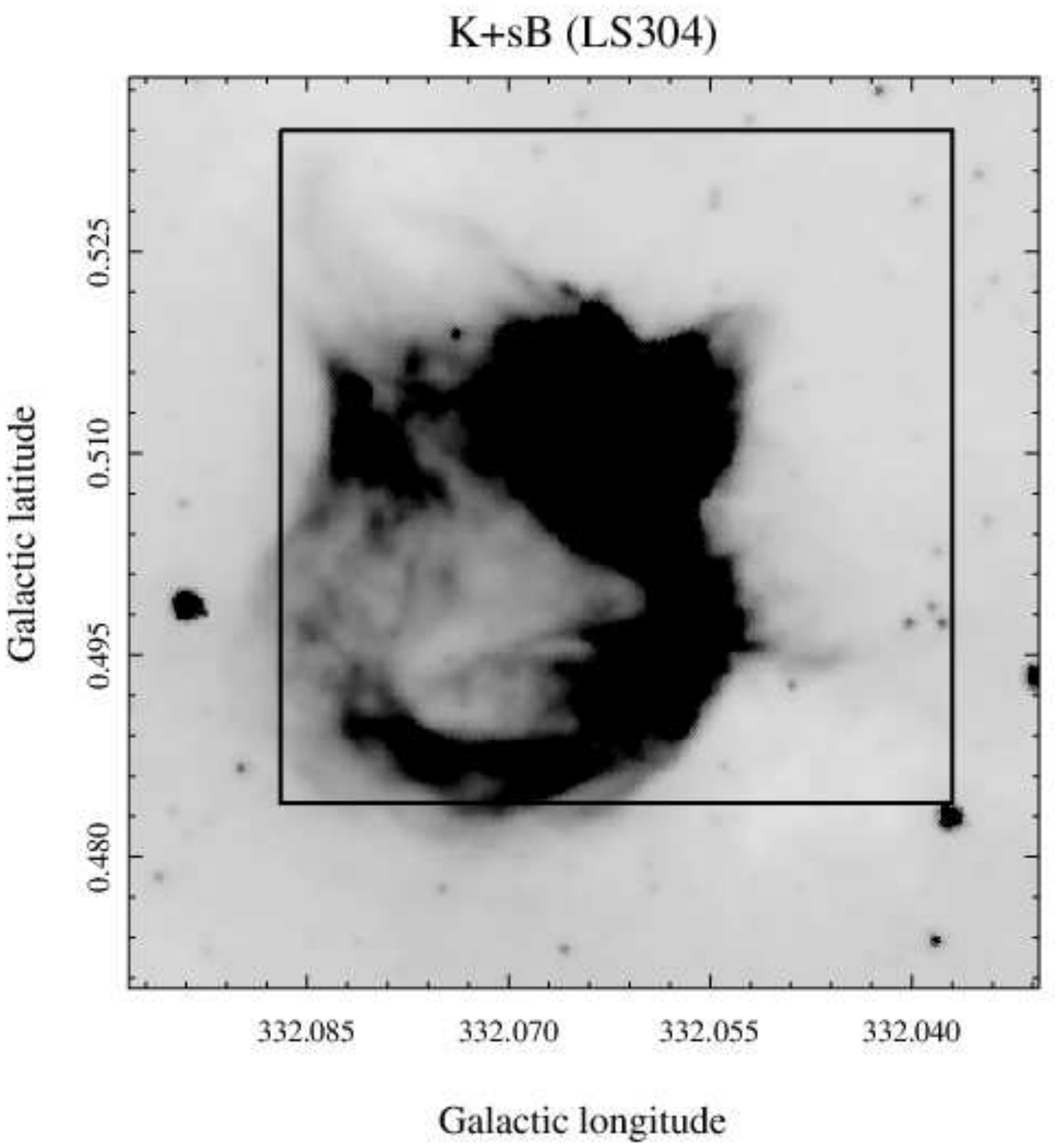}
\includegraphics[width=45mm]{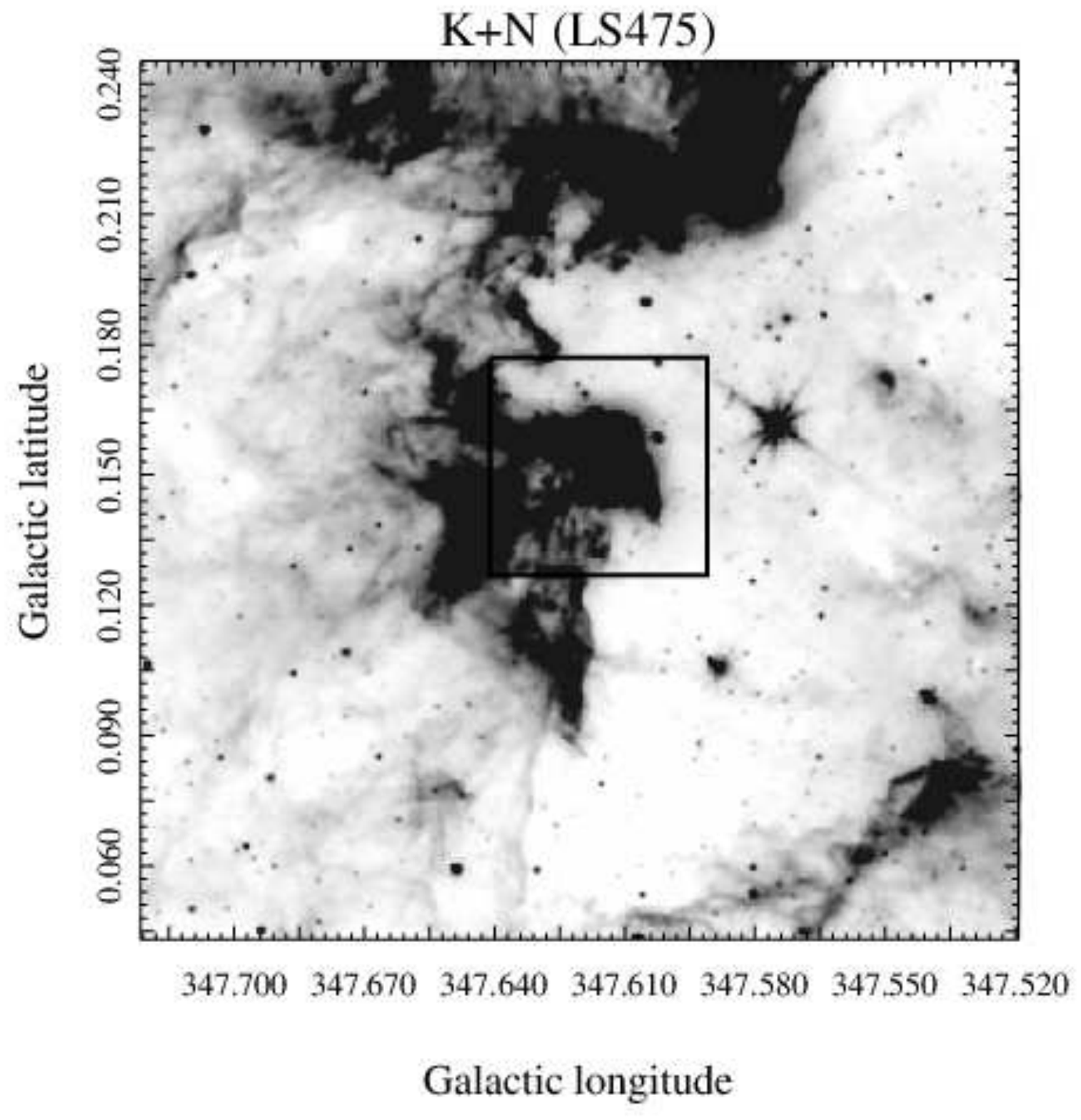}
\caption{Examples of different cluster morphologies based on images in the
  $8\,\mu$m band obtained with the Infrared Array Camera (IRAC) on board the
  Spizer Space Telescope. Boxes inside each panel correspond to the cluster
  candidate charts  showed in Figure~\ref{vvv-laserena_examples}.}  
   \label{spizer_examples}
\end{figure*}

In view of the importance of the rich morphological information that stands
out in the Spitzer 8\,$\mu$m images, we visually inspected the nebular
structures in the GLIMPSE images at the positions of the star cluster
candidates, which are described in column 11 of Table~\ref{tab1} 
The GLIMPSE survey extend approximately from Galactic latitude $-1\fdg2$ to
$1\fdg2$, and then only 441 cluster candidates are projected in that area,
with all of them detectable in some way (as point sources or extended 
structures) in the $8\,\mu$m images. 
Spitzer $8\,\mu$m point sources are the main feature for 80 cluster
candidates.  
Very compact nebulosity associated with point sources is present in 292
cluster candidates at that wavelength range. 
These kinds of sources are described as a knot. 
In many cases (156), these point sources and/or knots are also associated with 
extended irregular nebulosities (marked as nebula), and there are 89
clusters where this nebulosity is bubble-shaped, which were identified as
bubbles or small bubbles.  
Taking the three different nebular emission structures cataloged into account,
we count 361 cluster candidates with some kind of nebular emission 
associated at $8\,\mu$m, that means 82\% of the sample in the area of GLIMPSE
images.  
Figure\,\ref{spizer_examples} depicts some examples of clusters candidates as
seen in GLIMPSE $8\,\mu$m images. 
The square plotted inside each image corresponds to the size of the
VVV image charts shown in Figure\,\ref{vvv-laserena_examples}. 

The presence of knots and nebulosities in the Spitzer images suggests that
the clusters could be extremely young.
Taking this possibility into account, we explore the spatial correlation
between the cluster candidates and other independent tracers of ongoing 
star formation, such as masers, young stellar objects (YSO), 
extended green objects (EGO), and outflows.  

Astrophysical maser emissions are proven signatures of ongoing star formation,
in particular methanol masers \citep[cf.][]{2006ApJ...638..241E,
  2013MNRAS.435..524B}, although we must consider that water or hydroxyl
masers have also been detected in in star-forming regions, as well 
as in evolved stars or supernova remnants. 
Therefore, the presence of these maser types in the vicinity of a cluster
could be an excellent signspot that this cluster has ongoing star formation. 
We checked the presence of masers in the surrounding area of candidates using
the SIMBAD database and maser catalogs stored in Vizier.
The idea of our search is to highlight those cluster candidates where maser
emission is detected and then to provide a reference for the type of maser
present.
Table 2 lists the detected maser type, the source name (Column 6), and a
bibliographic reference for a catalog or discovery paper (Column 7).

\begin{figure}
   \centering
\includegraphics[width=8cm]{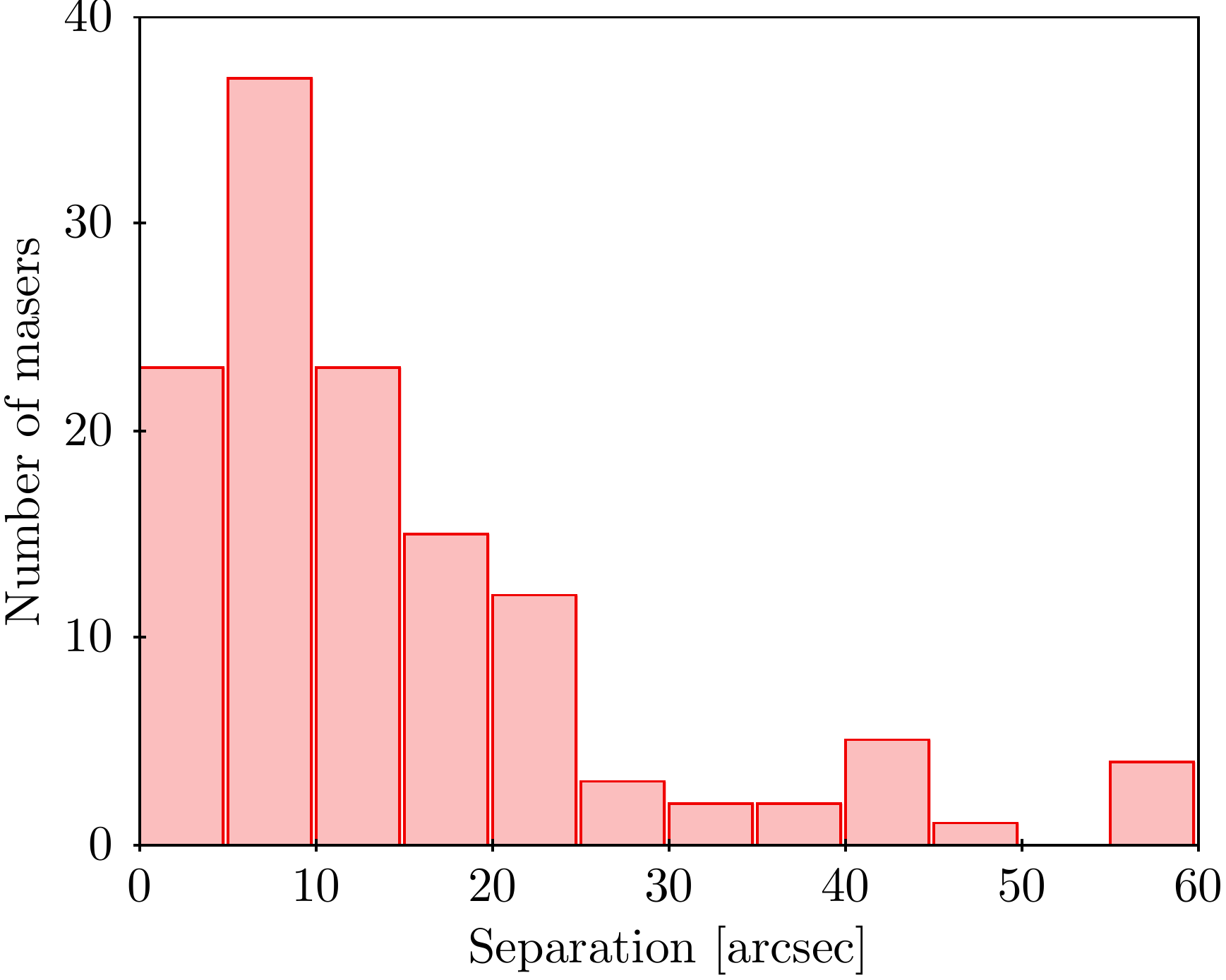}
  \caption{Histogram of the number of masers associated with the new VVV -- La
    Serena star cluster candidates.} 
   \label{xmatch-masers-clusters}
\end{figure}

We have found that 128 cluster candidates are associated with maser emissions
in a radius of $60\arcsec$, with 50\% of detections (63) within the
$10.5\arcsec$ circle (Figure~\ref{xmatch-masers-clusters}).
They correspond to 104 methanol masers, 72 water masers, and 55 hydroxil
masers. 
Detections of methanol and water masers are boosted for the efficiency of
dedicated surveys carried out by \cite{2010MNRAS.404.1029C}, 
\cite{2011MNRAS.417.1964C}, and \cite{2012MNRAS.420.3108G} for the
methanol and \cite{2011MNRAS.416.1764W} and \cite{2014MNRAS.442.2240W} for
water. 
It is interesting to note that 39 cluster candidates simultaneously include
the detection of the three type of masers (methanol, hydroxil, and water),
putting particular emphasis on these cluster candidates.
\cite{2010MNRAS.406.1487B} point out that the simultaneous detection of the
three type of masers perhaps indicate a special moment in the evolution of
newly formed stars; as these authors mention, the methanol masers are the
first species to develop and also the first species to disappear, 
whereas both water and hydroxil masers are more persistent species.  

\begin{figure}
   \centering
\includegraphics[width=8cm]{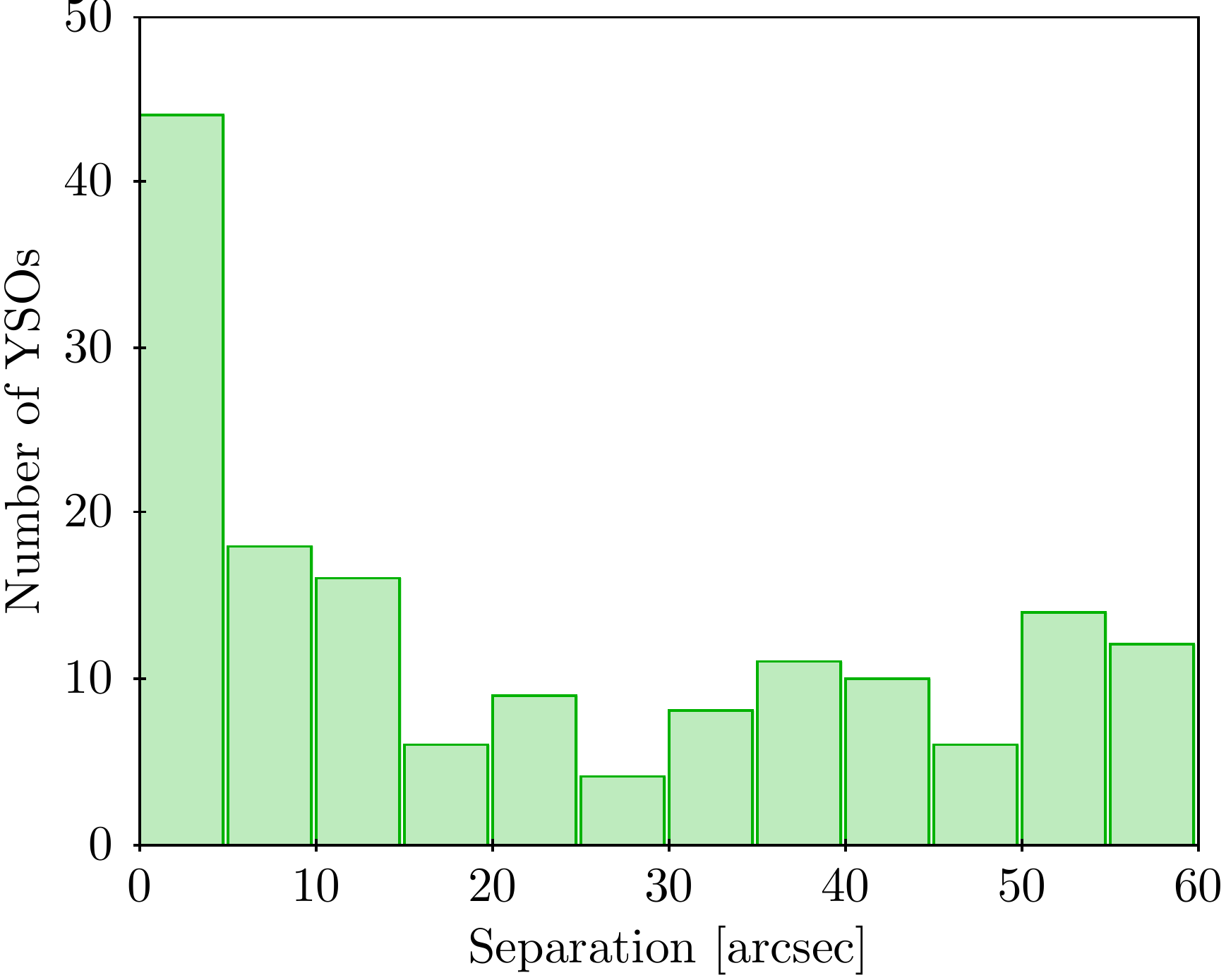}
  \caption{Histogram of the number of YSO candidates associated with the new
    VVV -- La Serena star cluster candidates.} 
   \label{xmatch-yso-clusters}
\end{figure}

YSOs are the primary tracers for star-forming regions.
Catalogs of YSOs and YSO candidates in the area of Galactic disk covered
by VVV survey include \cite{2007A&A...476.1019M} and
\cite{2014MNRAS.440.1213Y} (derived from MSX data), \cite{2008AJ....136.2413R}
(derived from Spitzer/GLIMPSE data), and \cite{1996A&AS..115..285C} (derived
from IRAS data).
Again, we performed a spatial cross-matching of star cluster candidates and
YSO candidates extracted from the mentioned catalogs, using a $60\arcsec$
radius.
The procedure results in the detection of 207 YSOs candidates in the area of
158 cluster candidates ($32\%$), which are $50\%$ of clusters with YSO
candidates inside the $15\arcsec$ circle, (Figure\,\ref{xmatch-yso-clusters}). 

\begin{figure}
   \centering
\includegraphics[width=9cm]{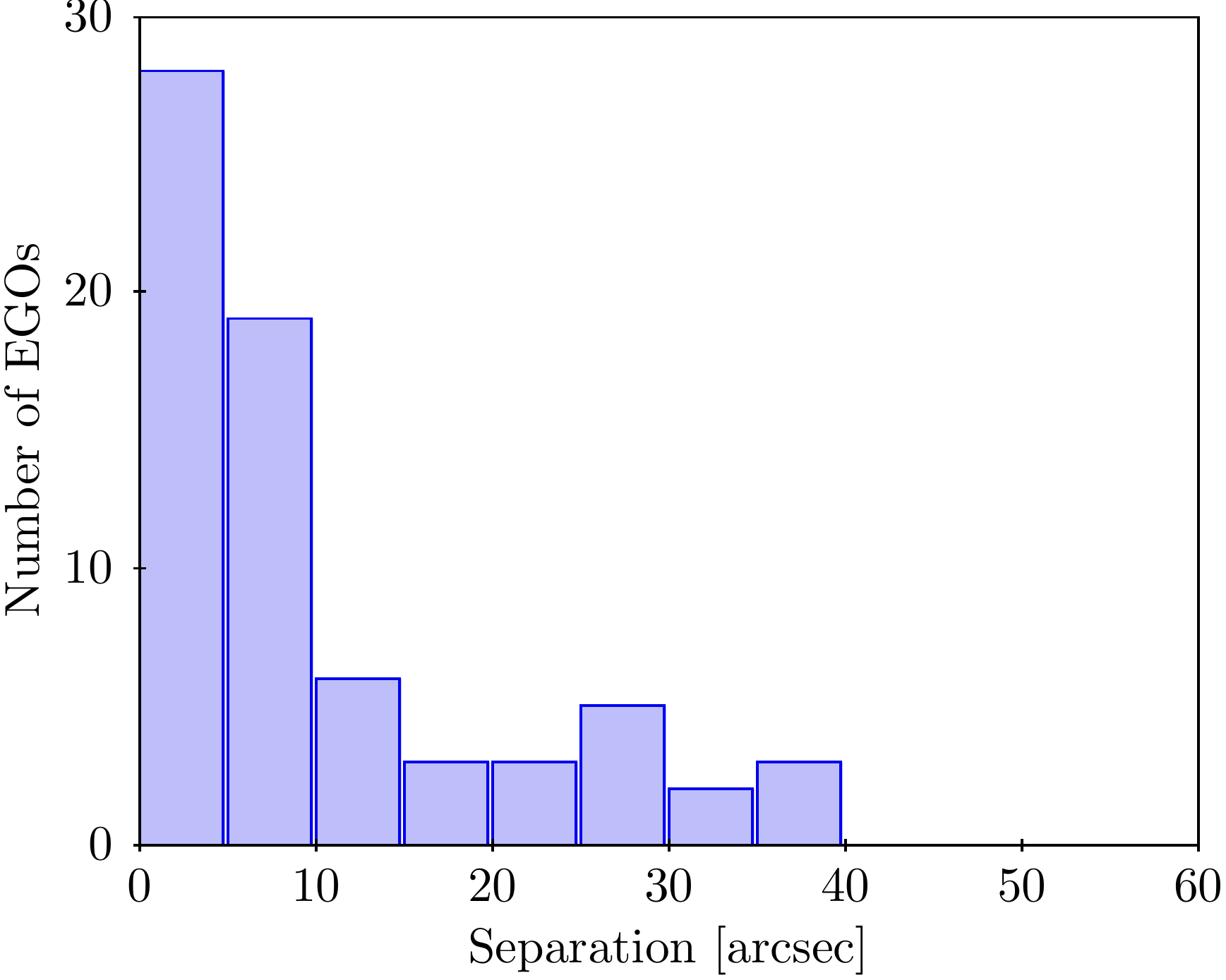}
  \caption{Histogram of the number of EGO and outflow candidates associated
    with the new VVV -- La Serena star cluster candidates.} 
   \label{xmatch-ego-clusters}
\end{figure}

Other interesting infrared tracers of ongoing star formation are outflows, 
extended green objects \citep[EGOs;][]{2008AJ....136.2391C}, and green fuzzies
\citep{2009ApJS..181..360C}.
These last two types of objects have acquired their names from the remarkable
$4.5\,\mu$m emission that is green in GLIMPSE color-composite images. 
EGOs and green fuzzies have been demostrated to be associated with young
high-mass star forming regions \citep{2009MNRAS.396.1603C,
  2011ApJS..196....9C}. 
We searched for catalogs about outflows and outflows candidates in ViZier, and
found two additional catalogs. 
A search for ionized outflows toward high-mass YSO was carried out by
\cite{2012ApJ...753...51G} through multifrequency radio contnuum observations
using the Australia Telescope Compact Array, and a catalog of molecular
outflows recompiled by \cite{2004A&A...426..503W}. 
The cross-matching between La Serena cluster catalog with outflows and EGOs
and catalogs brings 73 clusters candidates associated with 90 outflows
and EGO sources inside the circle with $60\arcsec$ radius, where $50\%$ are
inside $6.2\arcsec$ (see Figure\,\ref{xmatch-ego-clusters}). 

\section{Summary}

We report the discovery of 493 new near-infrared stellar cluster candidates in
a portion of the Galactic disk in the fourth quadrant using the ESO/Chile
public survey VISTA variables in the V\'{\i}a L\'actea.
The analysis of the spatial distribution shows that the new clusters are 
concentrated near the Galactic plane and feature conspicuous local maxima, 
corresponding to large star-forming complexes, such as \object{G305},
\object{RCW~95}, \object{RCW~106}.
Almost a half of the sample show a very compact appearance (250 candidates).  

The majority of the discovered cluster candidates (456, $92\%$) are
associated with dark infrared nebulosities, with 310 clusters directly 
associated to 552 cataloged dark clouds.

IRAS point sources are associated with $59\%$ of cluster sample, while for the
case of MSX point sources, the cross-matching gives $88\%$ of the new cluster
associated to a mid-infrared source. 
In view of the importance of identificating mid-IR counterparts for the vast
majority of new cluster candidates, we visually inspected GLIMPSE $8\,\mu$m
images. 
The morphological richness of the images is noteworthy. 
In 292 cluster candidates show knotty morphology at $8\,\mu$m, being 361
objects associated with some kind of extended emission nebula at this
wavelength regime. 
These morphologies suggest that most of the cluster candidates could be
extremely young.  

This scenario of extreme youth is reinforced by the cross-matching of the new
cluster candidates against masers, YSO candidates, outflows, and EGOs.  
In the first instance, 128 clusters candidates have associated maser emission,
with 104 of them methanol masers, which means excellent candidates for ongoing
massive star formation. 
Special mention is deserved by 39 cluster candidates that have associated
three maser types: methanol, water and hydroxil.
For YSO candidates and outflows, 158 ($32\%$) and 73 ($15\%$) of new clusters
candidates are closer than $60\arcsec$ from them, respectively.
In particular, there is a special set of sixteen cluster candidates that
present a  clear signpost of star-forming activity having associated
simultaneosly dark nebulae, young   stellar objects, extended green objects,
and maser emission. The cluster candidates are La~Serena numbers: 
045, 115, 119, 161, 227, 
228, 241, 283, 287, 316,
319, 391, 393, 410, 418, and
441.

Again, these findings suggest that a substantial proportion of the discovered
clusters are very young and have ongoing star formation. 
This scenario is consistent with the common features of the majority of the
clusters discovered independently by \cite{2011A&A...532A.131B} and
\cite{2014A&A...562A.115S}, using the same VVV DR1 images.

Up to now,  the vast majority of star cluster candidates discovered on the
basis of VVV images are quite compact and generally surrounded by bright or
dark nebulosities.  
With this work, we think that we are getting a the deeper level of
discovery using the VVV survey in the Galactic disk. 
But much remains to explore in relation to the population of slightly evolved
young clusters, where the detectable low-mass stars have left the accretion
phase.     
After this phase, the clusters have dissipated much of the molecular cloud in
which they formed, and some of the most massive and notable members have
evolved or have been ejected and may show a lower concentration.
These clusters are very difficult to separate the contaminating population of
the disk of the galaxy. 
Further opportunities to squeeze the VVV survey is to exploit the stack of
Ks-band images obtained during the current synoptic stage of the project. 
At present, VVV is gathering dozens of Ks-band images that can be the raw
material for pursuiting further immersed and fainter stellar clusters. 

\begin{acknowledgements}
We thank the anonymous referee for the very useful suggestions that helped us
improve this work. 
RHB acknowledges support by FONDECYT Regular project No. 1120668. 
ARL is grateful for financial support from ALMA-CONICYT Fund, and 
Direcci\'on de Investigaci\'on of Universidad de La Serena through project
DIULS REGULAR PR13144.  
JLNC also acknowledges financial support from the ALMA-CONICYT postdoctoral
program No. 31120026. 
MS acknowledges support from ESO-Chile Joint Committee and ''Becas Chile de Postdoctorado en el Extranjero'' project No. 74150088.
VF acknowledges support from the ESO-Chile Joint Committee and DIULS. 
MH acknowledges the support of the Basal PFB-06 and the Milenio Milky Way
Nucleus. 
We gratefully acknowledge use of data from the ESO Public Survey programme ID
179.B-2002, taken with the VISTA telescope, data products from the Cambridge
Astronomical Survey Unit, and funding from the FONDAP Center for Astrophysics
15010003, the BASAL CATA Center for Astrophysics and Associated Technologies
PFB-06, and the Millennium Institute of Astrophysics MAS. 
DM acknowledge financial support from CONICYT by Proyecto FONDECYT Regular
No. 1130196.  
R.K.S. acknowledges support from CNPq/Brazil through projects 310636/2013-2
and 481468/2013-7.
This research has made use of SIMBAD, VizieR, and ''Aladin sky atlas'' operated
and developed at CDS, Strasbourg Observatory, France. 
Also, this research has made use of NASA's Astrophysics Data System.

\end{acknowledgements}

\listofobjects


\onecolumn
\centering

\begin{landscape}
\begin{table}
\caption[La Serena star cluster candidates.]{La Serena star cluster candidates.} 
\label{tab1} 
\begin{tabular}{lllllllllll}

\hline\hline

Cluster ID &  R.A.$_{J2000}$ & Dec$_{J2000}$ &  $\ell$ &  $b$ & Tile & IRAS source & MSX source & Type &Size&  $8\mu$m morphology \\ 

\hline

La Serena 001 	   &    11:39:13	& $-63:29:04$   & 294.9726 	&   	$-1.7292$ 	&	d001   &	IRAS 11368-6312		&	G294.9725-01.7290   	&     C 	&       M    	&       -  	 	\\
La Serena 002 	   &    11:39:22	& $-63:28:11$   & 294.9896 	&   	$-1.7188$ 	&	d001   &	--             		&	G294.9895-01.7186   	&     C 	&       M    	&       -        	\\
La Serena 003 	   &    11:40:28	& $-63:27:58$   & 295.1026 	&   	$-1.6779$ 	&	d001   &	IRAS 11380-6311		&	G295.1026-01.6778   	&     C 	&       S    	&       -        	\\
La Serena 004 	   &    11:42:09	& $-62:20:07$   & 294.9846 	&   	$-0.5365$ 	&	d039   &	--             		&	G294.9845-00.5363   	&     O 	&       M    	&       P+N      	\\
La Serena 005 	   &    11:43:25	& $-62:25:43$   & 295.1523 	&   	$-0.5891$ 	&	d039   &	--             		&	G295.1523-00.5890   	&     O 	&       S    	&       -        	\\
La Serena 006 	   &    11:43:28	& $-62:27:20$   & 295.1652 	&   	$-0.6072$ 	&	d039   &	--             		&	G295.1651-00.6070   	&     O 	&       L    	&       P        	\\
La Serena 007 	   &    11:43:38	& $-62:25:17$   & 295.1750 	&   	$-0.5738$ 	&	d039   &	IRAS 11412-6208		&	G295.1750-00.5737   	&     O 	&       L    	&       K+N      	\\
\hline
\end{tabular}
\tablefoot{Only a small portion of the data is provided here, the full table is only available in electronic form at the CDS.}
\end{table}
\end{landscape}

\begin{landscape}
\begin{table}
\caption[La Serena star cluster candidates.]{Association with different
  astrophysical sources.}
\label{tab2}
\tiny
\begin{tabular} {lllllll}

\hline \hline

Cluster Id      & Solin et al. (2014) 	& Dark nebulae  &YSO candidates & EGO
and outflows  & Masers & Maser references \\ 
\hline
La Serena 001 &	       	        & 																			                                                              &                                                                                                                                                                                                                                                                                                                                                                              &																			      & \vtop{\hbox{\strut CH3OH [VES2000] 113651.54-631209.4  }\hbox{\strut H2O [BSG89] 294.97-1.73 }}											   & \vtop{\hbox{\strut 2000MNRAS.317..315V  }\hbox{\strut 1989A\&AS...77..465B}}                                                                       \\
\hline
\end{tabular}
\tablefoot{Only a small portion of the data is provided here, the full table is only available in electronic form at the CDS.}
\end{table}
\end{landscape}

\begin{table}
\caption[La Serena star cluster candidates.]{Special notes and comments for
  each cluster candidate.}
\label{tab3} 
\begin{tabular} {ll}

\hline 
\hline

Cluster ID      & Notes		  \\ 
\hline
La Serena 001 	   &1.6' from cluster VVV-CL 005. 1.8' from IC2944 and 1' from IC2948                                                  \\
La Serena 002 	   &1.4' from cluster VVV-CL 101                                                                                            \\
La Serena 003 	   &4' from IC 2944                                                                                             \\
La Serena 004 	   &                                                                                                            \\
La Serena 005 	   &In HII region [WMG70] 295.2-00.6                                                                            \\
La Serena 006 	   &Projected on open cluster Stock 14                                                                               \\
La Serena 007 	   &5' from open cluster Stock 14, and 1.3' from SNR G295.2-00.6                                                        \\
\hline
\end{tabular}
\tablefoot{Only a small portion of the data is provided here, the full table is only available in electronic form at the CDS.}
\end{table}

 \end{document}